\documentclass[
    preprintnumbers, 
    showpacs, 
    amsmath, 
    showkeys, 
    amssymb, 
    aps, 
    superscriptaddress, 
    onecolumn, 
    longbibliography, 
    letterpaper,
]{revtex4-2}

\usepackage{xcolor}
\usepackage{graphicx}
\usepackage{float}
\usepackage{adjustbox}
\usepackage{amsmath}
\usepackage{amsfonts}
\usepackage{mathrsfs}
\usepackage{braket}
\usepackage{booktabs, multirow}
\usepackage{tabularx}
\usepackage{longtable}
\usepackage{array}

\newcolumntype{C}{>{\centering\arraybackslash}X}
\newcolumntype{C}{>{$}c<{$}} % (｡•̀ᴗ-)✧ 单元格自动数学模式

\usepackage{placeins}
\usepackage{dblfloatfix}
\usepackage{tikz}
\usepackage{tikz-feynman}
\tikzfeynmanset{compat=1.1.0}
\usepackage{orcidlink}
\usepackage{hyperref}
\hypersetup{
    colorlinks=true,
    linkcolor=red,
    citecolor=blue,
}
\usepackage{bm}

\tikzfeynmanset{
    every diagram/.style={scale=1.15, transform shape},
    every edge/.append style={line width=0.7pt},
    every particle/.style={font=\scriptsize},
    every edge label/.style={font=\scriptsize, inner sep=1pt},
    boson/.style={decorate, decoration={snake, amplitude=1.1pt, segment length=3.0pt}, draw},
    scalar/.style={dashed, draw},
}

\newcommand{\lab}[1]{}

\newcommand{\ci}[1]{}

\newcommand{\bqa}{\begin{eqnarray}}
\newcommand{\eqa}{\end{eqnarray}}
\newcommand{\bqe}{\begin{equation}}
\newcommand{\eqe}{\end{equation}}
\newcommand{\bay}[1]{\left(\begin{array}{#1}}
\newcommand{\eay}{\end{array}\right)}

\begin{document}

% ============================================================
% 作者信息
% ============================================================
\author{Xiaoxuan Lin\orcidlink{0009-0005-6568-8453}}
\email{x-x.lin@stumail.hbu.edu.cn}
\affiliation{College of Physics Science and Technology, Hebei University, Baoding 071002, China}
\affiliation{Institute of Modern Physics, Chinese Academy of Sciences, Lanzhou 730000, China}

\author{Qian Wu}
\email{qwu@nju.edu.cn}
\affiliation{School of Physics, Nanjing University, Nanjing, Jiangsu 210093, China}
\affiliation{Institute of Modern Physics, Chinese Academy of Sciences, Lanzhou 730000, China}
\affiliation{Nuclear Many-Body Theory Laboratory, RIKEN Nishina Center, RIKEN, Wako 351-0198, Japan}

\author{Wei Kou\orcidlink{0000-0002-4152-2150}}
\email{kouwei@impcas.ac.cn}
\affiliation{Institute of Modern Physics, Chinese Academy of Sciences, Lanzhou 730000, China}
\affiliation{School of Nuclear Science and Technology, University of Chinese Academy of Sciences, Beijing 100049, China}
\affiliation{Southern Center for Nuclear Science Theory (SCNT), Institute of Modern Physics, Chinese Academy of Sciences, Huizhou 516000, Guangdong Province, China}
\affiliation{State Key Laboratory of Heavy Ion Science and Technology, Institute of Modern Physics, Chinese Academy of Sciences, Lanzhou 730000, China}

\author{Xurong Chen}
\email{xchen@impcas.ac.cn (Corresponding Author)}
\affiliation{Institute of Modern Physics, Chinese Academy of Sciences, Lanzhou 730000, China}
\affiliation{School of Nuclear Science and Technology, University of Chinese Academy of Sciences, Beijing 100049, China}
\affiliation{Southern Center for Nuclear Science Theory (SCNT), Institute of Modern Physics, Chinese Academy of Sciences, Huizhou 516000, Guangdong Province, China}
\affiliation{State Key Laboratory of Heavy Ion Science and Technology, Institute of Modern Physics, Chinese Academy of Sciences, Lanzhou 730000, China}
% ============================================================
% 标题和摘要
% ============================================================
\title{Probing a Fifth Force in Muonic Atoms through Lamb Shifts and Hyperfine Structure}

% ═══════════════════════════════════════════════════════════════════
% ✨ ABSTRACT — 相干耦合描述 (ﾉ◕ヮ◕)ﾉ
% ═══════════════════════════════════════════════════════════════════
\begin{abstract}
Motivated by the ATOMKI anomalies observed in $^8$Be and $^4$He transitions,
we present a systematic survey of $X17$-induced Lamb shifts and hyperfine
splittings in muonic atoms with stable nuclei up to $Z\le 15$. The
muon--nucleus bound-state problem is solved within the Gaussian Expansion
Method using a unified Hamiltonian that includes the standard electromagnetic
baseline together with vector- and pseudoscalar-$X17$ exchange. A central
feature of this work is that the Lamb-shift and hyperfine sectors are treated
with different nuclear couplings: the spin-independent Lamb shift is described
by a coherent vector muon--nucleus interaction, while the spin-dependent
hyperfine sector is constructed isotope by isotope from shell-model spin
fractions. Benchmark muon couplings are chosen in accordance with the 2025
Muon $g\!-\!2$ result. We find a clear complementarity between mediator hypotheses. The vector
Lamb-shift signal grows strongly toward heavier nuclei, while in the hyperfine
sector the vector scenario preferentially enhances odd-$N$ nuclei and the
pseudoscalar scenario favors odd-$Z$ systems. To compare different systems on a
common experimental footing, we introduce the signal-to-precision ratio. Among
systems with existing precision benchmarks, the most promising near-term probes
of the vector Lamb-shift channel are $\mu d$, $\mu^3\mathrm{He}^+$, and
$\mu^4\mathrm{He}^+$. For future spectroscopy, the largest absolute vector
Lamb-shift signal is predicted in $\mu^{31}\mathrm{P}$, while the leading
$1S_{1/2}$ hyperfine targets are $\mu^{29}\mathrm{Si}$ for the vector scenario
and $\mu^{31}\mathrm{P}$ for the pseudoscalar scenario.  The dominant theoretical uncertainty arises from the Schmidt-model treatment of
nuclear spin content, indicating that beyond-Schmidt nuclear-structure input
will be the key ingredient for quantitative follow-up studies of the leading
hyperfine targets.
\end{abstract}
\maketitle
% ============================================================
% 正文
% ============================================================
\section{Introduction}
\label{sec:intro}

The anomalous internal pair creation reported by the ATOMKI collaboration in
excited transitions of $^8$Be~\cite{Krasznahorkay2016PRL} and
$^4$He~\cite{Be4_Krasz_2019} has motivated sustained interest in a light boson
with a mass near $17$~MeV. If such a particle exists, it would signal a new
force beyond the Standard Model and would have broad implications for
light-mediator phenomenology and precision tests of fundamental interactions.
Among the proposed explanations, the protophobic vector scenario remains one of
the most widely discussed possibilities~\cite{Feng2016PRL}, while pseudoscalar
and axial-vector alternatives have also been explored
extensively~\cite{Ellwanger2016JHEP,Kozaczuk2017PRD}. A key question is how to
construct observables that are sensitive not only to the existence of such a
mediator, but also to its coupling pattern and spin-parity structure.

Muonic atoms provide a particularly favorable setting for such tests. Because
the muon is about 200 times heavier than the electron, its Bohr orbit is
compressed to the nuclear scale, greatly enhancing sensitivity to short-range
Yukawa-type interactions relative to ordinary atoms. The success of
high-precision muonic spectroscopy in the proton-radius puzzle demonstrated that
tiny level shifts in muonic systems can be measured with remarkable
accuracy~\cite{Antognini2013Science}, while more recent measurements in the
helium sector have shown that sub-meV sensitivity is achievable in
practice~\cite{Schuhmann2025Science}. For a mediator with
$m_X \simeq 16.7~\mathrm{MeV}$, the corresponding range
$m_X^{-1}\approx 12~\mathrm{fm}$ overlaps closely with the region most strongly
probed by low-lying muonic wave functions in light and medium-light nuclei.
Muonic atoms are therefore natural laboratories for searching for short-range
MeV-scale forces.

Most previous discussions of possible $X17$ effects in muonic atoms have
focused on only a few light systems, especially muonic hydrogen and
deuterium~\cite{Jentschura2020PRA,Martynenko2008PAN,Krutov2011EPJD}. While such
systems are theoretically clean and experimentally important, they do not by
themselves reveal the broader nuclear systematics expected from a new
short-range interaction. In particular, different observables probe different
combinations of hadronic couplings. For a vector mediator, the spin-independent
$2S_{1/2}$--$2P_{1/2}$ Lamb shift is governed by a coherent muon--nucleus
interaction and therefore depends on the total nuclear vector charge summed over
all nucleons. Hyperfine splittings, by contrast, are spin dependent and probe
the proton and neutron spin content of each nucleus. The optimal targets for
Lamb-shift and hyperfine measurements therefore need not coincide, and a
meaningful survey must treat the two sectors differently at the nuclear level.

In this work, we present a unified study of $X17$-induced Lamb shifts and
hyperfine splittings in muonic atoms with stable nuclei up to $Z\le 15$. Our
analysis makes three main advances. First, we perform a nucleus-by-nucleus
survey across the light and medium-light region rather than restricting
attention to a few benchmark systems. Second, we construct the two observables
with different nuclear inputs: the vector Lamb shift is described through the
coherent muon--nucleus coupling proportional to $Zh_p'+Nh_n'$, whereas the
hyperfine sector is built isotope by isotope from proton and neutron spin
fractions. Third, we connect the predicted signals directly to experimental
reach by introducing the signal-to-precision ratio
\[
\mathcal{R}\equiv |\Delta E_{X17}|/\delta E,
\]
where $\delta E$ denotes the achieved or projected precision for the relevant
transition. This quantity provides a simple way to compare systems with very
different absolute energy scales on common experimental footing.

Several robust patterns emerge from the present study. The vector Lamb-shift
channel is coherently enhanced and grows strongly toward heavier nuclei. In the
hyperfine sector, the vector and pseudoscalar hypotheses exhibit complementary
target selectivity because they weight neutron and proton spin content
differently: vector exchange preferentially enhances odd-$N$ nuclei, whereas
pseudoscalar exchange favors odd-$Z$ systems. This observable-dependent
separation makes comparative muonic spectroscopy a promising tool not only for
constraining a possible fifth force, but also for discriminating among mediator
hypotheses if a signal is observed.

Our calculations are performed within the Gaussian Expansion
Method~\cite{Hiyama2003}, using a unified Hamiltonian that includes the standard
electromagnetic baseline together with vector- and pseudoscalar-$X17$ exchange.
Nuclear masses, charge radii, magnetic moments, and related inputs are taken
from evaluated databases~\cite{AME2020,Stone2005,Angeli2013}, while the nuclear
spin structure is treated at the independent-particle shell-model
(Schmidt-model) level~\cite{KraneBook,BohrMottelson1998}. On the leptonic side,
we adopt benchmark muon couplings guided by the updated Muon $g\!-\!2$
situation. Benchmark comparisons with published results for light muonic atoms
show that the electromagnetic baseline is sufficient for target identification,
while the dominant theoretical uncertainty in the hyperfine sector arises from
the Schmidt-model treatment of nuclear spin fractions rather than from the
atomic calculation itself.

The remainder of this paper is organized as follows.
Section~\ref{sec:theory} presents the theoretical framework, including the
Hamiltonian, coupling choices, nuclear spin input, and validation of the
numerical implementation.
Section~\ref{sec:results} gives the predicted Lamb-shift and hyperfine signals
for both mediator hypotheses and discusses their implications for experimental
target selection.
Section~\ref{sec:conclusion} summarizes the main conclusions and outlines the
prospects for ongoing and future muonic-atom spectroscopy programs.
Comprehensive numerical results are collected in
Appendices~\ref{app:lamb-vector}--\ref{app:hfs-ps}.
% ═══════════════════════════════════════════════════════════════════
% ✿ Feynman图 ✿
% ═══════════════════════════════════════════════════════════════════
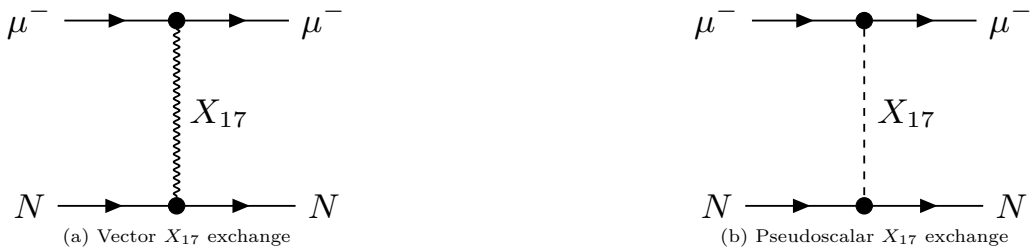
\begin{figure}[t]
\centering
\begin{minipage}[t]{0.49\linewidth}
\centering
\begin{tikzpicture}[scale=1.45, transform shape]
\begin{feynman}
  \vertex (i1) at (-1.35,  0.85) {\(\mu^-\)};
  \vertex (f1) at ( 1.35,  0.85) {\(\mu^-\)};
  \vertex (i2) at (-1.35, -0.85) {\(N\)};
  \vertex (f2) at ( 1.35, -0.85) {\(N\)};

  \vertex[dot] (a) at ( 0.00,  0.85) {};
  \vertex[dot] (b) at ( 0.00, -0.85) {};

  \diagram*{
    (i1) -- [fermion] (a) -- [fermion] (f1),
    (i2) -- [fermion] (b) -- [fermion] (f2),
    (a)  -- [boson, edge label=\(X_{17}\)] (b),
  };
\end{feynman}
\end{tikzpicture}

\vspace{-2mm}
{\scriptsize (a) Vector \(X_{17}\) exchange}
\end{minipage}
\hfill
\begin{minipage}[t]{0.49\linewidth}
\centering
\begin{tikzpicture}[scale=1.45, transform shape]
\begin{feynman}
  \vertex (i1) at (-1.35,  0.85) {\(\mu^-\)};
  \vertex (f1) at ( 1.35,  0.85) {\(\mu^-\)};
  \vertex (i2) at (-1.35, -0.85) {\(N\)};
  \vertex (f2) at ( 1.35, -0.85) {\(N\)};

  \vertex[dot] (a) at ( 0.00,  0.85) {};
  \vertex[dot] (b) at ( 0.00, -0.85) {};

  \diagram*{
    (i1) -- [fermion] (a) -- [fermion] (f1),
    (i2) -- [fermion] (b) -- [fermion] (f2),
    (a)  -- [scalar, edge label=\(X_{17}\)] (b),
  };
\end{feynman}
\end{tikzpicture}

\vspace{-2mm}
{\scriptsize (b) Pseudoscalar \(X_{17}\) exchange}
\end{minipage}
\caption{Schematic illustration of muon--nucleon interactions mediated by a
light boson \(X_{17}\). Panel (a) shows vector exchange, which contributes to
both the Lamb shift and hyperfine structure. Panel (b) shows pseudoscalar
exchange, which contributes only to the hyperfine sector at leading order.}
\label{fig:x17_exchange_hfs}
\end{figure}

\section{Theoretical framework}
\label{sec:theory}

\subsection{The System Hamiltonian}

% ── 两体系统: Lamb shift 用相干核耦合, HFS 用 spin-fraction 加权 ~~(ˊ˘ˋ)~~
We treat the problem as a muon--nucleus two-body system with Hamiltonian
\begin{equation}
H \;=\; H_0 \;+\; H_{\rm fs} \;+\; H_F \;+\; H^{X17}.
\label{eq:H_total}
\end{equation}
The atomic calculation is formulated entirely in terms of the relative coordinate
between the muon and the center-of-mass of the nucleus.  The underlying
new-physics interaction is defined at the muon--nucleon level and then
mapped onto muon--nucleus operators through two distinct procedures, depending on
the observable.  For the hyperfine sector, the spin-dependent muon--nucleon
interaction is mapped via the nuclear spin fractions $\Delta_{p,n}$ introduced in
Sec.~\ref{sec:nuclear-structure}, yielding spin-fraction-weighted effective
couplings.  For the spin-independent vector Lamb shift, the nucleon-level
couplings are instead summed coherently over all $Z$ protons and $N$ neutrons,
as described below and in Sec.~\ref{sec:lamb}.  This two-step procedure ensures
that the nuclear-structure information appropriate to each observable is cleanly
separated from the atomic-physics calculation.

The leading Schr\"{o}dinger Hamiltonian contains the kinetic energy with reduced mass
and the point Coulomb potential,
\begin{equation}
H_0 \;=\; \frac{\mathbf p^{\,2}}{2m_r}\;-\;\frac{Z\alpha}{r},
\qquad 
m_r \equiv \frac{m_\mu m_N}{m_\mu+m_N}.
\label{eq:H0}
\end{equation}

Following the standard Breit--Pauli expansion used in muonic atoms, the
relativistic, Darwin/finite-size, and leptonic spin--orbit contributions are
organized as Ref.~\cite{Martynenko2008PAN}
% ── 记号统一: σ_f → σ_μ ~~✧~~
\begin{align}
H_{\rm fs}
&= -\,\frac{\mathbf p^{\,4}}{8m_\mu^{3}}
-\,\frac{\mathbf p^{\,4}}{8m_{N}^{3}}
+ \frac{\pi Z\alpha}{2}
\!\left(\frac{4\langle r_{N}^{2}\rangle}{3}
+\frac{1}{m_\mu^{2}}+\frac{1}{m_{N}^{2}}\right)
\delta^{3}(\mathbf r) \nonumber\\
&\quad
+\;\frac{Z\alpha}{4m_\mu^{2} r^{3}}
\left(1+\frac{2m_\mu}{ m_{N}}\right)
\mathbf L\!\cdot\!\boldsymbol{\sigma}_\mu
\;+\; V_{\rm Ueh}(r),
\label{eq:H_fs}
\end{align}
where $\langle r_N^2\rangle$ is the mean-square nuclear charge radius.

In the present work, the Uehling vacuum-polarization potential enters only as a
background QED effect: its role is to provide the correct order of magnitude for
the standard Lamb-shift contribution against which we compare the much smaller
$X17$-induced shifts.  Since our primary focus is on the new-physics correction
from $X17$, and not on a high-precision re-evaluation of the vacuum-polarization
effect itself, it is sufficient for our purposes to employ a simple point-nucleus
model for the Uehling potential.

The Uehling vacuum-polarization potential was first derived in Ref.~\cite{Uehling:1935}
in the form
\begin{equation}
V_{\rm Ueh}(r)
= -\frac{Z\alpha}{r}\,\frac{\alpha}{\pi}
\int_0^1 (1-u^2)\,du
\int_1^\infty
\exp\!\left[
-\frac{2 r z}{\lambda_0 \sqrt{1-u^2}}
\right]
\frac{dz}{z} ,
\label{eq:Uehling_Uehling1935}
\end{equation}
where $\lambda_0 = \hbar/(m_e c)$ is the Compton wavelength of the electron.

% ── 记号统一: m_f → m_μ, σ_f → σ_μ ~~(｡•̀ᴗ-)✧~~
The Fermi Hamiltonian $H_F$ (see Eq.~(10) of Ref.~\cite{Jentschura2006PRA}) reads
\begin{equation}
\label{eq:H_F}
H_F \;=\; \frac{g_N\,\alpha}{m_\mu\,m_N}
\left[
\frac{\pi}{3}\;\boldsymbol{\sigma}_\mu\!\cdot\!\boldsymbol{\sigma}_N\,\delta^{(3)}(\mathbf r)
\;+\;
\frac{3\,(\boldsymbol{\sigma}_\mu\!\cdot\!\mathbf r)(\boldsymbol{\sigma}_N\!\cdot\!\mathbf r) - r^2\,\boldsymbol{\sigma}_\mu\!\cdot\!\boldsymbol{\sigma}_N}{8\,r^5}
\;+\;
\frac{\boldsymbol{\sigma}_N\!\cdot\!\mathbf L}{4\,r^3}
\right] .
\end{equation}

% ── ★ 核心修改: Lamb shift 用相干 Yukawa 势, HFS 用 spin-fraction (◕‿◕✿) ──
For a vector $X17$, the $00$ component of the propagator generates a leading,
spin-independent Yukawa interaction.  At the muon--nucleus level, coherently
summing over all $Z$ protons and $N$ neutrons yields
\begin{equation}
V_X(r) \;=\; \frac{h'_\mu}{4\pi}\,(Z h'_p + N h'_n)\,\frac{e^{-m_X r}}{r},
\label{eq:VX_vector}
\end{equation}
which governs the $2S_{1/2}$--$2P_{1/2}$ Lamb shift discussed in
Sec.~\ref{sec:lamb}.  Because Eq.~\eqref{eq:VX_vector} sums the vector charge
over the entire nuclear bulk, it is distinct from the spin-fraction-weighted
prescription used in the hyperfine sector (Sec.~\ref{sec:nuclear-structure}),
which retains only the spin-carrying valence nucleon(s).
The induced hyperfine Hamiltonian (vector exchange) is
\begin{align}
H_{\rm HFS,V}
&=\frac{h'_\mu h'^{(\rm eff)}_N}{16\pi\,m_\mu m_N}\Bigg[
-\frac{8\pi}{3}\,\delta^{(3)}(\mathbf r)\,\boldsymbol\sigma_\mu\!\cdot\!\boldsymbol\sigma_N
- m_X^{2}\,
\frac{\boldsymbol\sigma_\mu\!\cdot\!\mathbf r\;\boldsymbol\sigma_N\!\cdot\!\mathbf r - r^{2}\,\boldsymbol\sigma_\mu\!\cdot\!\boldsymbol\sigma_N}{r^{3}}\,e^{-m_X r} \nonumber\\
&\hspace{2.1cm}
-\,(1+m_X r)\,
\frac{3\,\boldsymbol\sigma_\mu\!\cdot\!\mathbf r\;\boldsymbol\sigma_N\!\cdot\!\mathbf r - r^{2}\,\boldsymbol\sigma_\mu\!\cdot\!\boldsymbol\sigma_N}{r^{5}}\,e^{-m_X r}
-\left(2+\frac{m_\mu}{m_N}\right)
\frac{(1+m_X r)}{r^{3}}\,\boldsymbol\sigma_N\!\cdot\!\mathbf L\,e^{-m_X r}
\Bigg],
\label{eq:H_HFS_vector}
\end{align}
where $h'^{(\rm eff)}_N = h_p'\Delta_p + h_n'\Delta_n$ is the spin-fraction-weighted
effective coupling defined in Eq.~\eqref{eq:heff}.

% ── pseudoscalar: only spin-dependent, no centroid shift ~~(✿◠‿◠)~~
For a pseudoscalar $X17$ with pointlike $i\gamma_5$ couplings, the tree-level
exchange induces a purely spin-dependent potential.  Because its matrix elements
vanish in the spin-averaged centroids of the $nS$ and $nP$ levels, this potential
does not generate an independent shift of the level centroids comparable to the
vector Lamb shift; rather, it redistributes the energy within each hyperfine
multiplet and thereby modifies the observable hyperfine splittings.  The
corresponding hyperfine Hamiltonian is
\begin{align}
H_{\mathrm{HFS},A}
&= \frac{h_\mu\,h^{(\rm eff)}_N}{16\pi\,m_\mu\,m_N}\Bigg[
\frac{4\pi}{3}\,\delta^{(3)}(\mathbf r)\,
\boldsymbol{\sigma}_\mu\!\cdot\!\boldsymbol{\sigma}_N
\;-\;
m_X^{2}\,
\frac{(\boldsymbol{\sigma}_\mu\!\cdot\!\mathbf r)\,
    (\boldsymbol{\sigma}_N\!\cdot\!\mathbf r)}{r^{3}}\,
e^{-m_X r} \nonumber\\
&\hspace{3.0cm}
+\,(1+m_X r)\,
\frac{3(\boldsymbol{\sigma}_\mu\!\cdot\!\mathbf r)\,
    (\boldsymbol{\sigma}_N\!\cdot\!\mathbf r)
    - r^{2}\,\boldsymbol{\sigma}_\mu\!\cdot\!\boldsymbol{\sigma}_N}{r^{5}}\,
e^{-m_X r}
\Bigg],
\label{eq:H_HFS_pseudoscalar}
\end{align}
where $h^{(\rm eff)}_N = h_p\Delta_p + h_n\Delta_n$.

% ── S 态角平均化简 ~~(ˊ˘ˋ)~~
On spherically symmetric $S$-state wave functions one may use the angular averages
\begin{equation}
(\boldsymbol{\sigma}_\mu\!\cdot\!\mathbf r)\,(\boldsymbol{\sigma}_N\!\cdot\!\mathbf r)
\;\longrightarrow\; \frac{1}{3}\,r^{2}\,\boldsymbol{\sigma}_\mu\!\cdot\!\boldsymbol{\sigma}_N,
\qquad
\boldsymbol{\sigma}_N\!\cdot\!\mathbf L \;\longrightarrow\; 0,
\end{equation}
so that
\begin{equation}
H_{\rm HFS,V}\to
-\,\frac{h'_\mu h'^{(\rm eff)}_N}{24\pi\,m_\mu m_N}\;
\boldsymbol{\sigma}_\mu\!\cdot\!\boldsymbol{\sigma}_N
\left[\,4\pi\,\delta^{(3)}(\mathbf r)\;-\;\frac{m_X^{2}}{r}\,e^{-m_X r}\right].
\label{eq:HHFSV_reduced}
\end{equation}
\begin{equation}
H_{\rm HFS,A}\to
\;\;\frac{h_\mu h^{(\rm eff)}_N}{48\pi\,m_\mu m_N}\;
\boldsymbol{\sigma}_\mu\!\cdot\!\boldsymbol{\sigma}_N
\left[\,4\pi\,\delta^{(3)}(\mathbf r)\;-\;\frac{m_X^{2}}{r}\,e^{-m_X r}\right].
\label{eq:HHFSA_reduced}
\end{equation}
\begin{equation}
H_F\to
\frac{g_N\,\alpha}{m_\mu m_N}\,
\frac{\pi}{3}\,\boldsymbol{\sigma}_\mu\!\cdot\!\boldsymbol{\sigma}_N\,
\delta^{(3)}(\mathbf r).
\label{eq:H_Fermi}
\end{equation}

% ── P 态处理 ~~✧~~
For $P$ states it is convenient to separate angles and radius via $\mathbf r=r\,\hat{\mathbf r}$:
\begin{equation}
(\boldsymbol{\sigma}_\mu\!\cdot\!\mathbf r)(\boldsymbol{\sigma}_N\!\cdot\!\mathbf r)
= r^{2}(\boldsymbol{\sigma}_\mu\!\cdot\!\hat{\mathbf r})(\boldsymbol{\sigma}_N\!\cdot\!\hat{\mathbf r})
= \frac{r^{2}}{3}\!\left[\boldsymbol{\sigma}_\mu\!\cdot\!\boldsymbol{\sigma}_N+S_{12}\right],
\quad
S_{12}=3(\boldsymbol{\sigma}_\mu\!\cdot\!\hat{\mathbf r})(\boldsymbol{\sigma}_N\!\cdot\!\hat{\mathbf r})
-\boldsymbol{\sigma}_\mu\!\cdot\!\boldsymbol{\sigma}_N .
\label{eq:sigmar-reduction}
\end{equation}

Matrix elements are evaluated in the coupled basis
$\big|\big((L\,s_\mu)J\,S_N\big)\,F m_F\big\rangle$.
Projectors onto fixed $J$ satisfy
\begin{equation}
\mathbb P_J\,\mathbf L\,\mathbb P_J=\alpha_J\,\mathbf J,\qquad
\mathbb P_J\,\boldsymbol{\sigma}_\mu\,\mathbb P_J=2(1-\alpha_J)\,\mathbf J,
\qquad
\alpha_J=\frac{J(J+1)+L(L+1)-s_\mu(s_\mu+1)}{2\,J(J+1)}.
\label{eq:projectors}
\end{equation}
With $\boldsymbol{\sigma}_N\equiv2\mathbf S_N$ and
\begin{equation}
\Delta\equiv F(F+1)-J(J+1)-S_N(S_N+1),\qquad
\mathbf J\!\cdot\!\boldsymbol{\sigma}_N=\Delta,
\label{eq:Delta-def}
\end{equation}
the $F$-resolved expectation values take the universal form (at fixed $L,s_\mu,J$):
\begin{equation}
\langle \mathbf L\!\cdot\!\boldsymbol{\sigma}_N\rangle=\alpha_J\,\Delta,\qquad
\langle \boldsymbol{\sigma}_\mu\!\cdot\!\boldsymbol{\sigma}_N\rangle=2(1-\alpha_J)\,\Delta,\qquad
\langle \mathbf L\!\cdot\!\boldsymbol{\sigma}_\mu\rangle=J(J{+}1)-L(L{+}1)-s_\mu(s_\mu{+}1).
\label{eq:universal-P}
\end{equation}

For $1S_{1/2}$ one has $L=0$, $J=\tfrac12$, hence $\alpha_{1/2}=0$ and
\[
\langle\mathbf L\!\cdot\!\boldsymbol{\sigma}_\mu\rangle=0,\quad
\langle\mathbf L\!\cdot\!\boldsymbol{\sigma}_N\rangle=0,\quad
\langle \boldsymbol{\sigma}_\mu\!\cdot\!\boldsymbol{\sigma}_N\rangle=2\,\Delta.
\]

For $2P_{1/2}$ ($L=1$, $s_\mu=\tfrac12$, $J=\tfrac12$) one finds $\alpha_{1/2}=4/3$, thus
\[
\langle \mathbf L\!\cdot\!\boldsymbol{\sigma}_N\rangle=\tfrac{4}{3}\Delta,\qquad
\langle \boldsymbol{\sigma}_\mu\!\cdot\!\boldsymbol{\sigma}_N\rangle=-\tfrac{2}{3}\Delta .
\]

Treating $S_{12}$ as a scalar coupling of rank-2 tensors yields, for $2P_{1/2}$,
\begin{equation}
\big\langle (\boldsymbol{\sigma}_\mu\!\cdot\!\hat{\mathbf r})
(\boldsymbol{\sigma}_N\!\cdot\!\hat{\mathbf r}) \big\rangle
= -\frac{14}{27}\,\Delta,
\qquad
\big\langle S_{12}\big\rangle = -\frac{8}{9}\,\Delta.
\label{eq:RR-S12-2P12}
\end{equation}

These obey the exact identity
\begin{equation}
\frac{1}{8}\,\langle S_{12}\rangle + \frac{1}{4}\,\langle \boldsymbol{\sigma}_N\!\cdot\!\mathbf L\rangle
= -\frac{1}{3}\,\langle \boldsymbol{\sigma}_\mu\!\cdot\!\boldsymbol{\sigma}_N\rangle,
\label{eq:combo-identity}
\end{equation}
which is a convenient diagnostic for implementations.  For $S$ waves, the tensor
reduces to
$(\boldsymbol{\sigma}_\mu\!\cdot\!\mathbf r)(\boldsymbol{\sigma}_N\!\cdot\!\mathbf r)\to \frac{r^{2}}{3}\,\boldsymbol{\sigma}_\mu\!\cdot\!\boldsymbol{\sigma}_N$,
and the spin-orbit terms vanish.

% ── 角动量期望值表格 ~~(｡•̀ᴗ-)~~
\begin{table}[H]
\centering
\caption{$\langle \mathbf L\!\cdot\!\boldsymbol{\sigma}_\mu\rangle$ for $1S_{1/2}$ and $2P_{1/2}$.
Here $\boldsymbol{\sigma}_\mu$ is the muon spin operator (Pauli convention).}
\label{tab:LSf}
\setlength{\tabcolsep}{10pt}
\begin{tabular}{cc}
\toprule
State & $\langle \mathbf L\!\cdot\!\boldsymbol{\sigma}_\mu\rangle$ \\
\midrule
$1S_{1/2}$  & $0$  \\
$2P_{1/2}$  & $-2$ \\
\bottomrule
\end{tabular}
\end{table}

\begin{table}[H]
\centering
\caption{$\langle \mathbf L\!\cdot\!\boldsymbol\sigma_N\rangle$ and $\langle \boldsymbol\sigma_\mu\!\cdot\!\boldsymbol\sigma_N\rangle$ for $1S_{1/2}$ (Pauli normalization).}
\label{tab:spin_1S}
\setlength{\tabcolsep}{10pt}
\renewcommand{\arraystretch}{1.2}
\begin{tabular}{@{} c c c c @{}}
\toprule
\multirow{2}{*}{$S_N$} & \multirow{2}{*}{$\langle \mathbf L\!\cdot\!\boldsymbol\sigma_N\rangle$}
& \multicolumn{2}{c}{$\langle \boldsymbol\sigma_\mu\!\cdot\!\boldsymbol\sigma_N\rangle$} \\ 
\cmidrule(lr){3-4}
& & $F=S_N-\frac{1}{2}$ & $F=S_N+\frac{1}{2}$ \\
\midrule
$\tfrac12$ & $0$ & $-3$ & $+1$ \\
$1$        & $0$ & $-4$ & $+2$ \\
$\tfrac32$ & $0$ & $-5$ & $+3$ \\
$\tfrac52$ & $0$ & $-7$ & $+5$ \\
\bottomrule
\end{tabular}
\end{table}

\begin{table*}[t]
\centering
\caption{\(\langle \mathbf{L}\!\cdot\!\boldsymbol{\sigma}_N\rangle\) and \(\langle \boldsymbol{\sigma}_\mu\!\cdot\!\boldsymbol{\sigma}_N\rangle\) for \(2P_{1/2}\) (Pauli normalization).}
\label{tab:spin_2P12}
\setlength{\tabcolsep}{8pt}
\renewcommand{\arraystretch}{1.2}
\begin{tabular}{@{} c cc cc @{}}
\toprule
\multirow{2}{*}{$S_N$} &
\multicolumn{2}{c}{$\langle \mathbf{L}\!\cdot\!\boldsymbol{\sigma}_N\rangle$} &
\multicolumn{2}{c}{$\langle \boldsymbol{\sigma}_\mu\!\cdot\!\boldsymbol{\sigma}_N\rangle$} \\
\cmidrule(lr){2-3}\cmidrule(lr){4-5}
& $F=S_N-\tfrac12$ & $F=S_N+\tfrac12$ & $F=S_N-\tfrac12$ & $F=S_N+\tfrac12$ \\
\midrule
$\tfrac12$ & $-2$            & $+\tfrac{2}{3}$ & $+1$            & $-\tfrac{1}{3}$ \\
$1$        & $-\tfrac{8}{3}$ & $+\tfrac{4}{3}$ & $+\tfrac{4}{3}$ & $-\tfrac{2}{3}$ \\
$\tfrac32$ & $-\tfrac{10}{3}$& $+2$            & $+\tfrac{5}{3}$ & $-1$            \\
$\tfrac52$ & $-\tfrac{14}{3}$& $+\tfrac{10}{3}$& $+\tfrac{7}{3}$ & $-\tfrac{5}{3}$ \\
\bottomrule
\end{tabular}
\end{table*}

\begin{table*}[t]
\centering
\caption{%
\(\big\langle(\boldsymbol{\sigma}_\mu\!\cdot\!\mathbf r)\,
(\boldsymbol{\sigma}_N\!\cdot\!\mathbf r)\big\rangle\) for \(2P_{1/2}\) 
\emph{(Pauli normalization; angle--radius separated)}.
We first evaluate the pure angular factor
\(\langle(\boldsymbol{\sigma}_\mu\!\cdot\!\hat{\mathbf r})(\boldsymbol{\sigma}_N\!\cdot\!\hat{\mathbf r})\rangle
= -\tfrac{14}{27}\,\Delta\) with 
\(\Delta=F(F{+}1)-J(J{+}1)-S_N(S_N{+}1)\), \(J=\tfrac12\), and then multiply by \(r^2\).}
\label{tab:sigr_sigr_2P12}
\setlength{\tabcolsep}{12pt}
\renewcommand{\arraystretch}{1.3}
\begin{tabular}{@{} c cc @{}}
\toprule
\multirow{2}{*}{$S_N$} &
\multicolumn{2}{c}{\(\big\langle(\boldsymbol{\sigma}_\mu\!\cdot\!\mathbf r)\,
    (\boldsymbol{\sigma}_N\!\cdot\!\mathbf r)\big\rangle\)} \\
\cmidrule(lr){2-3}
& $F=S_N-\tfrac{1}{2}$ & $F=S_N+\tfrac{1}{2}$ \\
\midrule
$\tfrac{1}{2}$ & $\tfrac{7}{9}\,r^{2}$   & $-\tfrac{7}{27}\,r^{2}$ \\
$1$            & $\tfrac{28}{27}\,r^{2}$ & $-\tfrac{14}{27}\,r^{2}$ \\
$\tfrac{3}{2}$ & $\tfrac{35}{27}\,r^{2}$ & $-\tfrac{7}{9}\,r^{2}$ \\
$\tfrac{5}{2}$ & $\tfrac{49}{27}\,r^{2}$ & $-\tfrac{35}{27}\,r^{2}$ \\
\bottomrule
\end{tabular}
\end{table*}

% ── GEM 实现 ~~(ˊ˘ˋ)~~
\subsection{GEM implementation}

We start from the Rayleigh--Ritz principle on the Hilbert space $\mathcal H$ of
square-integrable two-body wavefunctions. The stationary states minimize the
energy functional under the unit-norm constraint,
\begin{equation}
\mathcal F[\psi,\lambda]
=\langle\psi|\hat H|\psi\rangle-\lambda\big(\langle\psi|\psi\rangle-1\big),
\qquad
\delta\mathcal F=0,
\label{eq:RR}
\end{equation}
which yields the Euler--Lagrange condition $(\hat H-\lambda)|\psi\rangle=0$ with $\lambda=E$. 

For a central field it is natural to separate angles and solve, for each partial
wave $(L,m)$,
\begin{equation}
\Big[-\,\frac{\nabla^2}{2\,\mu}+V(r)\Big]\psi_{Lm}(\mathbf r)=E\,\psi_{Lm}(\mathbf r),
\qquad
\mu=\frac{m_\mu m_N}{m_\mu+m_N}.
\label{eq:radial}
\end{equation}

To make the constrained minimization explicit, expand
$|\psi\rangle=\sum_{j}c_j\,|\phi_j\rangle$ in an arbitrary (generally non-orthogonal)
set $\{|\phi_j\rangle\}\subset\mathcal H$. In coefficient space the functional
becomes the Rayleigh quotient
\begin{equation}
\mathcal F(\mathbf c,\lambda)=\mathbf c^\dagger H\,\mathbf c-\lambda\big(\mathbf c^\dagger N\,\mathbf c-1\big),
\qquad
H_{ij}=\langle\phi_i|\hat H|\phi_j\rangle,\;
N_{ij}=\langle\phi_i|\phi_j\rangle,
\label{eq:RR-coeff}
\end{equation}
and stationarity with respect to $\mathbf c^\ast$ gives the generalized Hermitian
eigenproblem
\begin{equation}
H\,\mathbf c=E\,N\,\mathbf c.
\label{eq:GEP-abstract}
\end{equation}

Since $N$ is positive-definite on the span of $\{|\phi_j\rangle\}$,
\eqref{eq:GEP-abstract} can be reduced to a standard Hermitian problem
$B^{-1}HB^{-\dagger}\,\mathbf y=E\,\mathbf y$ with the variational upper-bound
property and monotone convergence as the span is enlarged.

Within the Gaussian Expansion Method (GEM) \cite{Hiyama2003,Kamimura1988,SuzukiVarga1998},
we choose a radial basis that retains analytic integrability while densely sampling
short and long distances:
\begin{align}
\psi_{\ell m}(\mathbf r)
&= \sum_{n=1}^{n_{\max}} c_{n\ell}\,\phi^{G}_{n\ell m}(\mathbf r),
\label{eq:GEM-exp} \\[2pt]
\phi^{G}_{n\ell m}(\mathbf r)
&= \mathcal N_{n\ell}\; r^{\ell} e^{-\nu_n r^{2}}\, Y_{\ell m}(\hat{\mathbf r}),
\qquad
\mathcal N_{n\ell}
= \left[
\frac{2^{\ell+2}\,(2\nu_n)^{\ell+\frac{3}{2}}}
{\sqrt{\pi}\,(2\ell{+}1)!!}
\right]^{\!1/2},
\label{eq:Gauss-def} \\[2pt]
N_{n,n'}
&= \big\langle \phi^{G}_{n\ell m} \,\big|\, \phi^{G}_{n'\ell m} \big\rangle
= \left(
\frac{2\sqrt{\nu_n \nu_{n'}}}{\nu_n + \nu_{n'}}
\right)^{\ell+\frac{3}{2}}.
\label{eq:overlap}
\end{align}
with geometric placement of ranges to ensure near-uniform resolution in $\log r$,
\begin{equation}
\nu_n=\frac{1}{r_n^2},\qquad
r_n=r_{\min}\,a^{\,n-1},\qquad
a=\Big(\frac{r_{\max}}{r_{\min}}\Big)^{1/(n_{\max}-1)},
\quad n=1,\ldots,n_{\max}.
\label{eq:grid}
\end{equation}

Plugging \eqref{eq:GEM-exp} into \eqref{eq:radial} and using \eqref{eq:RR-coeff}
we obtain, for each $L$,
\begin{equation}
\sum_{n'=1}^{n_{\max}}\!\Big[T^{(L)}_{nn'}+V^{(L)}_{nn'}-E\,N^{(L)}_{nn'}\Big]\,c_{n'L}=0.
\label{eq:GEP-L}
\end{equation}

% ── X17 耦合 ~~✿~~
\subsection{Couplings to X17}
\label{sec:couplings-x17}

On the hadronic side we adopt two convenient benchmarks.  First, for a
\emph{vector} mediator we write
\begin{equation}
h_p'=\epsilon_p\,e,\qquad h_n'=\epsilon_n\,e,
\end{equation}
with nucleon charges built from quark charges,
\begin{equation}
\epsilon_p=2\epsilon_u+\epsilon_d,\qquad
\epsilon_n=\epsilon_u+2\epsilon_d.
\end{equation}

Motivated by fits to the $^{8}\mathrm{Be}$ anomaly we take the working values
\begin{equation}
\epsilon_p=8\times10^{-4},\qquad
\epsilon_n=\frac{1}{100}\,,
\end{equation}
which imply
\begin{align}
h_p'&=\epsilon_p\,e\simeq 2.42\times10^{-4},\\
h_n'&=\epsilon_n\,e\simeq 3.03\times10^{-3},
\end{align}
with $\epsilon_n$ controlled by $\epsilon_{u,d}$ as above~\cite{Feng2016PRL}.

Second, for a pseudoscalar mediator we take couplings proportional to fermion
masses,
\begin{equation}
h_p=\xi_p\,\frac{m_p}{v},\qquad h_n=\xi_n\,\frac{m_n}{v},\qquad v=246~\mathrm{GeV},
\end{equation}
which, using standard nucleon matrix elements (as in \cite{Jentschura2020PRA}),
give
\begin{align}
h_p &\simeq \frac{m_p}{v}\,\big(-0.40\,\xi_u-1.71\,\xi_d\big)
\approx\; -2.4\times10^{-3},\\[4pt]
h_n &\simeq \frac{m_n}{v}\,\big(-0.40\,\xi_u+0.85\,\xi_d\big)
\approx\; 5.1\times10^{-4}.
\end{align}

Turning to the lepton side, we consider both a vector and a pseudoscalar $X17$
that couple to the muon.  For the vector hypothesis we write
\begin{equation}
\mathcal{L}_{X,V} \;=\; -\,h'_\mu\,\bar\mu\gamma^\nu \mu\,X_\nu 
\end{equation}
and for the pseudoscalar hypothesis
\begin{equation}
\mathcal{L}_{X,A} \;=\; -\,h_\mu\,\bar\mu\, i\gamma_5 \mu\,A\,.
\end{equation}

At one loop these interactions shift the muon anomaly $a_\mu=(g-2)_\mu/2$.
For a mediator around $m_X\simeq 16.7~\mathrm{MeV}$, the standard one-loop
expressions read \cite{Leveille1978,LindnerPlatscherQueiroz2018}
\begin{align}
\Delta a_\mu^{(V)}
&= \frac{(h'_\mu)^2}{8\pi^2}\frac{m_\mu^2}{m_X^2}
\int_0^1\!\mathrm{d}x\,
\frac{x^2(2-x)}{(1-x)\!\left(1-\tfrac{m_\mu^2}{m_X^2}\right)+\tfrac{m_\mu^2}{m_X^2}x}
\;\simeq\; 8.64\times 10^{-3}\,(h'_\mu)^2, \label{eq:amuV}\\[2pt]
\Delta a_\mu^{(A)}
&= -\,\frac{h_\mu^{\,2}}{4\pi^2}\frac{m_\mu^2}{m_X^2}
\int_0^1\!\mathrm{d}x\,
\frac{x^3}{(1-x)\!\left(1-\tfrac{m_\mu^2}{m_X^2}\right)+\tfrac{m_\mu^2}{m_X^2}x}
\;\simeq\; -\,1.19\times 10^{-3}\,h_\mu^{\,2}.
\label{eq:amuA}
\end{align}

Thus a vector $X17$ \emph{raises} $a_\mu$ (positive sign), while a pseudoscalar
\emph{lowers} it (negative sign).

Using the latest experimental world average from the Muon $g\!-\!2$ experiment at
Fermilab~\citep{MuonG2_2025}, we take
\begin{equation}
a_\mu^{\exp}=116\,592\,070.5(14.5)\times10^{-11}.
\label{eq:amuexp}
\end{equation}

For the Standard-Model prediction we use the 2025 update by the Muon
$g\!-\!2$ Theory Initiative~\citep{TI_WP25},
\begin{equation}
a_\mu^{\mathrm{SM}}=116\,592\,033(62)\times10^{-11}.
\label{eq:amuSM}
\end{equation}

These imply
\begin{equation}
\Delta a_\mu \equiv a_\mu^{\exp}-a_\mu^{\mathrm{SM}}=3.8(6.3)\times 10^{-10}.
\label{eq:Deltaamu}
\end{equation}

It is important to note that this discrepancy represents only a $0.6\sigma$
deviation, largely due to the 2025 Theory Initiative's adoption of lattice-QCD
results for the hadronic vacuum polarization (HVP) contribution. This marks a
significant shift from the earlier $\sim 4\sigma$ tension reported by the 2020
Theory Initiative white paper, which relied primarily on dispersive evaluations.
While the theoretical situation remains under active investigation---with ongoing
efforts to reconcile the lattice and dispersive approaches---the current picture
suggests substantially reduced room for new-physics contributions to the muon
anomaly.

Nevertheless, for the purpose of this work, we adopt benchmark couplings that
remain consistent with the $\pm 6\sigma$ envelope around the measured discrepancy.
We denote by $\sigma_a$ the one-sigma width for $\Delta a_\mu$,
\begin{equation}
\sigma_a \equiv 6.3\times 10^{-10}.
\label{eq:sigmaa}
\end{equation}

Given the 2025 update of the Muon $g\!-\!2$ Theory Initiative, which adopts a
consolidated lattice-QCD average for the LO-HVP contribution, the current
difference corresponds to only a $\sim0.6\sigma$ effect and is therefore
statistically consistent with zero at the present precision.  In this situation,
our goal is not to provide a best-fit explanation of the muon anomaly, but rather
to define representative upper-scale benchmark couplings for assessing the
potential size of $X17$-induced atomic shifts and their systematic uncertainties.
We therefore allow $\Delta a_\mu$ to vary within a wide interval
$\Delta a_\mu\pm 6\sigma_a$, which avoids imposing a strong prior on the
unsettled HVP treatment and prevents overly restrictive benchmark couplings that
would trivially suppress the signal.  Since all $X17$-induced energy shifts considered here scale as $h_\mu^2$ (or $(h'_\mu)^2$), our results can be straightforwardly rescaled to
any narrower confidence interval if desired.

To define benchmark targets reflecting the current situation, in which the
experimental central value lies above the SM prediction by $+0.6\sigma$, we
choose
\begin{equation}
\Delta a_\mu^{(V)}=+0.6\,\sigma_a=+3.78\times10^{-10},\qquad
\Delta a_\mu^{(A)}=-(6.0-0.6)\,\sigma_a=-5.4\,\sigma_a=-3.402\times10^{-9}.
\label{eq:benchmarkDelta}
\end{equation}

Inserting these targets into Eqs.~\eqref{eq:amuV}--\eqref{eq:amuA}, the benchmark
muon-side couplings are
\begin{align}
h'_\mu &= 2.09\times10^{-4}, \label{eq:hmuV}\\
h_\mu  &= -1.69\times10^{-3}. \label{eq:hmuA}
\end{align}
These benchmark choices should not be interpreted as statistically preferred fits.
Rather, they serve as representative upper-scale couplings consistent with the
sign of the one-loop contributions and with the broadened $\pm 6\sigma$ range
adopted in this work.

% ── 核结构 ~~(ˊ˘ˋ)✧~~
\subsection{Nuclear structure and magnetic response for $Z\!\le\!15$ nuclei}
\label{sec:nuclear-structure}

In light nuclei ($Z\!\le\!15$) the independent-particle shell model provides a
quantitatively successful zeroth-order description of ground-state spins and
magnetic moments~\cite{Mayer1949,Jensen1955,BohrMottelson1998,KraneBook}. Pauli
blocking fills the lowest single-particle orbitals in time-reversed pairs, so their
spin contributions cancel and the net magnetic and spin-dependent response is
dominated by the unpaired valence nucleon(s) in the open
shell~\cite{NeugartNeyens2017,Stone2005}.

We parameterize the nucleon-side coupling of an $X17$ mediator to a given nucleus
through spin fractions $\Delta_{p,n}$ that encode the proton/neutron share of the
nuclear spin:
\begin{equation}
\Delta_{p,n}
\;\equiv\;
\frac{\big\langle \sum_{i\in p,n}\mathbf S_i \big\rangle\!\cdot\!\hat{\mathbf I}}
{\hbar\,I},
\qquad
\big\langle \sum_{i\in p,n}\mathbf S_i \big\rangle
= \alpha^{(S)}_{lj}\,\hbar\,\hat{\mathbf I}\,.
\label{eq:DeltaDef}
\end{equation}
Here $I$ and $\hat{\mathbf I}$ are the nuclear spin magnitude and its unit vector.
For a single Schmidt nucleon with orbital angular momentum $l$ and total angular
momentum $j = l\pm\tfrac{1}{2}$, the spin-projection coefficient is
\begin{equation}
\alpha^{(S)}_{lj}
\;=\;
\frac{j(j{+}1)+s(s{+}1)-l(l{+}1)}{2(j{+}1)}\,,\qquad s=\tfrac12.
\label{eq:alphaS}
\end{equation}
For even-even ($I{=}0$) nuclei, $\Delta_p=\Delta_n=0$ by definition.

When both a valence proton and a valence neutron contribute to the ground-state
spin (odd-odd nuclei), their angular momenta $j_p$ and $j_n$ couple to the total
$I$ in the weak-coupling limit.  The projection coefficients are
\begin{align}
C_p &= \frac{I(I{+}1)+j_p(j_p{+}1)-j_n(j_n{+}1)}{2\,I(I{+}1)},\qquad
C_n = \frac{I(I{+}1)+j_n(j_n{+}1)-j_p(j_p{+}1)}{2\,I(I{+}1)}, \label{eq:Cpn}\\[4pt]
\Delta_p &= \frac{C_p\,\alpha^{(S)}_{l_pj_p}}{I}\,,\qquad
\Delta_n = \frac{C_n\,\alpha^{(S)}_{l_nj_n}}{I}\,.
\label{eq:DeltasWeakCoupling}
\end{align}
These reproduce the standard additivity formulas used in nuclear magnetic-moment
systematics~\cite{NeugartNeyens2017,KraneBook,Stone2005}; for odd-$A$ nuclei the
expressions reduce to the single-Schmidt form.

Given $\Delta_{p,n}$, the effective nucleon-side couplings entering the
$X17$-induced hyperfine operators are
\begin{equation}
h_{N}^{\prime(\mathrm{eff})} = h_p'\,\Delta_p + h_n'\,\Delta_n\quad\text{(vector HFS)},
\qquad
h_{N}^{(\mathrm{eff})} = h_p\,\Delta_p + h_n\,\Delta_n\quad\text{(pseudoscalar HFS)}.
\label{eq:heff}
\end{equation}
These effective couplings enter exclusively the $X17$-induced hyperfine operators
of Eqs.~\eqref{eq:H_HFS_vector}--\eqref{eq:H_HFS_pseudoscalar}.  For the
vector-induced $2S_{1/2}$--$2P_{1/2}$ Lamb shift, the relevant muon--nucleus
coupling is instead the coherent sum
$Z h_p' + N h_n'$ appearing in Eq.~\eqref{eq:VX_vector}, which probes the total
spin-independent vector charge of the nucleus rather than its spin content.  The
numerical values of both $h_N^{\prime(\rm eff)}$ and $Zh_p'+Nh_n'$ for each
nuclide are listed in Appendix~\ref{app:inputs:spin}.

\begin{figure}[t]
\centering
\includegraphics[width=0.8\textwidth]{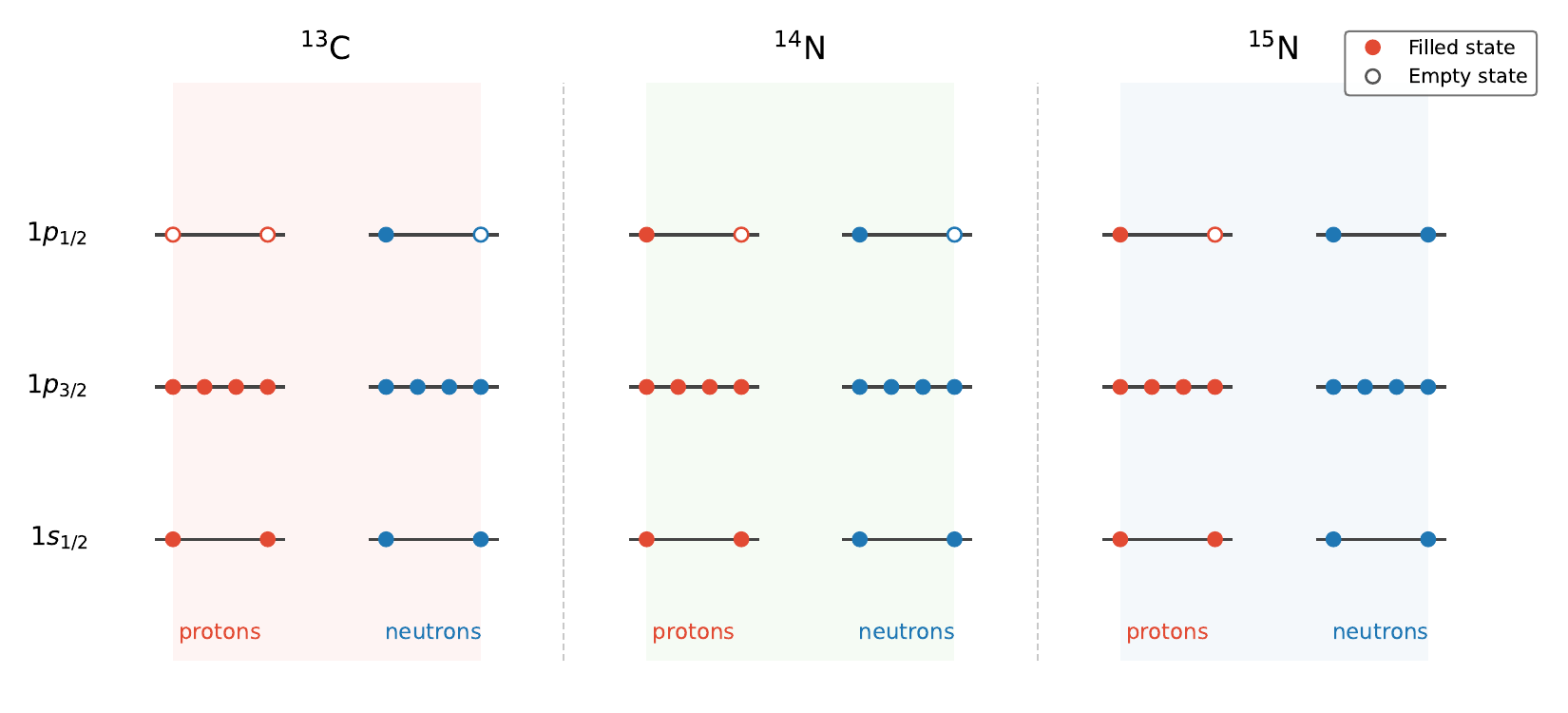}
\caption{%
Single-particle occupancies in the $p$ shell for $^{13}$C, $^{14}$N, and $^{15}$N.
The $1s_{1/2}$, $1p_{3/2}$, and $1p_{1/2}$ orbitals are shown; left (right) columns
correspond to proton (neutron) states.  Filled circles denote occupied states.}
\label{fig:p-shell-occupancies}
\end{figure}

We illustrate the procedure with three representative $p$-shell nuclei, whose
single-particle occupancies are shown in Fig.~\ref{fig:p-shell-occupancies}.
For $^{13}$C ($Z{=}6$, $N{=}7$, $I^\pi=\tfrac12^-$), the valence neutron occupies
$1p_{1/2}$; Eq.~\eqref{eq:alphaS} gives $\alpha^{(S)}_{p_{1/2}}=-\tfrac16$, so
that $\Delta_n = -\tfrac13$ and $\Delta_p=0$.
For $^{14}$N ($Z{=}7$, $N{=}7$, $I^\pi=1^+$), two valence nucleons with
$j_p=j_n=\tfrac12$ occupy $1p_{1/2}$; Eq.~\eqref{eq:Cpn} yields
$C_p=C_n=\tfrac12$, and with $I=1$ one finds $\Delta_p=\Delta_n=-\tfrac{1}{12}$.
For $^{15}$N ($Z{=}7$, $N{=}8$, $I^\pi=\tfrac12^-$), the neutron $p$ shell is
closed and the valence proton occupies $1p_{1/2}$, giving $\Delta_p=-\tfrac13$
and $\Delta_n=0$.

% ── Schmidt baseline 免责声明 ~~(˘ᵕ˘)~~
Configuration mixing and core polarization modify these Schmidt-model values at a
level that depends sensitively on the nuclear structure of each isotope and is not
uniformly well-constrained~\cite{Stone2005,BohrMottelson1998,NeugartNeyens2017}.
The orbital assignments and spin-projection coefficients adopted throughout this
work are used solely to define a consistent Schmidt-model baseline for the survey,
and should not be interpreted as a claim of a full shell-model description.  This
baseline is sufficient for the purpose of identifying the relative
$X17$-sensitivity trends across the nuclear chart; quantitative precision for the
highest-priority targets will require beyond-Schmidt calculations.

Unless otherwise noted, ground-state $J^\pi$ assignments and low-lying level
energies are taken from the evaluated databases TUNL~\cite{KelleyTUNL} and ENSDF
(accessed via NNDC NuDat3)~\cite{NNDC_NuDat3}; for isotonic-chain cross-checks
we consult the KAERI chart of nuclides~\cite{KAERI_Chart}. Shape information
motivating Nilsson assignments is supported by evaluated electric quadrupole
moments~\cite{Stone2016_Qmom} and the Bohr--Mottelson rotational-coupling
formalism~\cite{BohrMottelson1998}. For the $N{=}11$ isotones we adopt $^{21}$Ne
as prolate with $I^\pi=\tfrac{3}{2}^{+}$ and the first $5/2^{+}$ level at
$E\simeq351$\,keV~\cite{NNDC_NuDat3,KAERI_Chart,Stone2016_Qmom}.

The experimental nuclear $g$-factors are computed from evaluated magnetic moments
and spins via
\begin{equation}
g_N^{(\mathrm{exp})} \;=\; \frac{m_{\rm nuc}}{m_p}\,\frac{\mu_{\rm exp}}{\mu_N\,I}\,,
\label{eq:gN}
\end{equation}
where $\mu_{\rm exp}$ is expressed in nuclear magnetons
$\mu_N = e\hbar/(2m_pc)$ and the factor $m_{\rm nuc}/m_p$ enforces the convention
of Eq.~\eqref{eq:H_F}. Nuclear masses are from AME2020~\cite{AME2020}; magnetic
moments for the light nuclei $^{1}$H, $^{2}$H, $^{3}$H, $^{3}$He are taken from
PDG~\cite{PDG2024}, and those for stable nuclides with $Z\le15$ from Stone's
evaluated tables~\cite{Stone2005,Stone2016_Qmom}.

% ── Validation ~~(✿◠‿◠)~~
\subsection{Validation and systematic uncertainties}
\label{sec:validation}

Before presenting predictions for $X17$-induced shifts, we assess the robustness
of the theoretical framework against two sources of systematic error: higher-order
QED corrections to the wave function, and the point-nucleus approximation for the
Uehling potential.

\paragraph{Finite-size Uehling correction.}
The Uehling potential in Eq.~\eqref{eq:Uehling_Uehling1935} was derived for a
point nucleus. For nuclei with finite charge radius $R_{\rm ch}$, the proper
treatment convolves the point-nucleus potential with the nuclear charge
distribution~\cite{Fullerton1976,Borie1982}:
\begin{equation}
V_{\rm Ueh}^{\rm (fs)}(\mathbf{r})
= \int \frac{\rho_{\rm ch}(\mathbf{r}')}{Z}\,
  V_{\rm Ueh}^{\rm (pt)}(|\mathbf{r}-\mathbf{r}'|)\,\mathrm{d}^3\mathbf{r}',
\label{eq:V_Ueh_convolution}
\end{equation}
where $V_{\rm Ueh}^{\rm (pt)}(r)$ is given by Eq.~\eqref{eq:Uehling_Uehling1935}
and $\rho_{\rm ch}$ satisfies $\int\rho_{\rm ch}\,\mathrm{d}^3r'=Z$.  Adopting a
uniformly charged sphere of radius $R_{\rm ch}$,
\begin{equation}
\rho_{\rm ch}(r') =
\begin{cases}
\dfrac{3Z}{4\pi R_{\rm ch}^3}, & r' \le R_{\rm ch},\\[6pt]
0, & r' > R_{\rm ch},
\end{cases}
\label{eq:rho_uniform_sphere}
\end{equation}
the angular integration reduces to~\cite{Borie1982}
\begin{equation}
V_{\rm Ueh}^{\rm (fs)}(r)
= \frac{2\pi}{r}
  \int_0^\infty \frac{\rho_{\rm ch}(r')}{Z}\,r'^2\,\mathrm{d}r'
  \int_{|r-r'|}^{r+r'}
  V_{\rm Ueh}^{\rm (pt)}(s)\,s\,\mathrm{d}s,
\label{eq:V_Ueh_radial_convolution}
\end{equation}
which is evaluated numerically on the GEM radial mesh.

\begin{figure}[t]
\centering
\begin{minipage}[b]{0.48\textwidth}
\centering
\includegraphics[width=\textwidth]{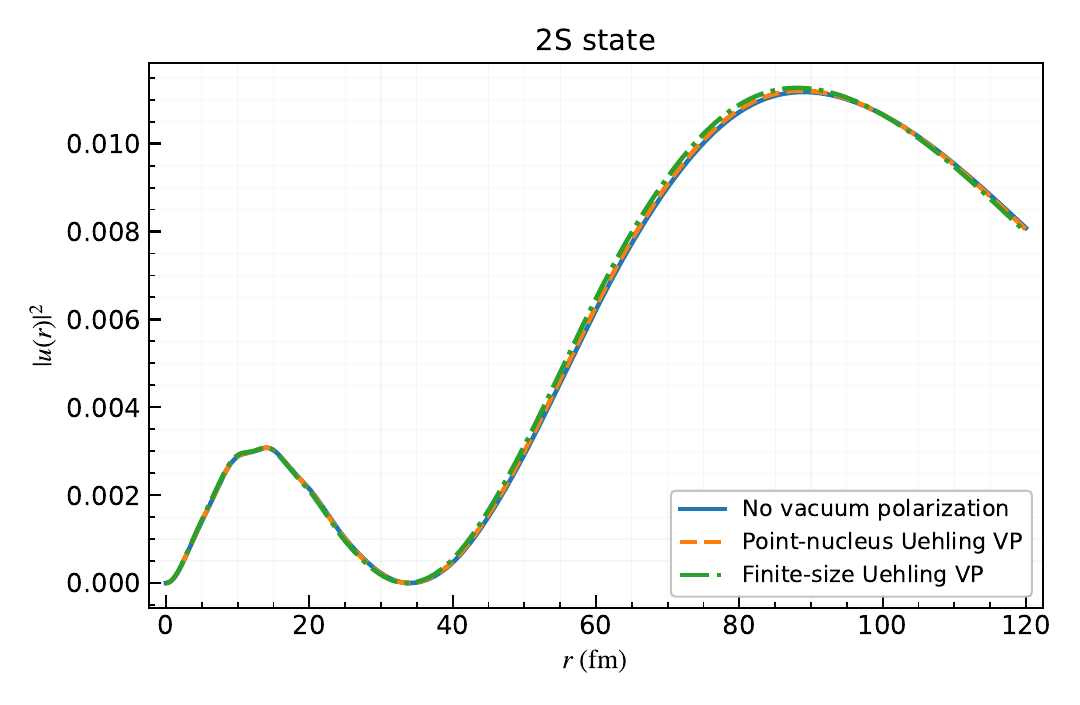}

\vspace{1mm}
{\small (a) $2S_{1/2}$ state}
\end{minipage}
\hfill
\begin{minipage}[b]{0.48\textwidth}
\centering
\includegraphics[width=\textwidth]{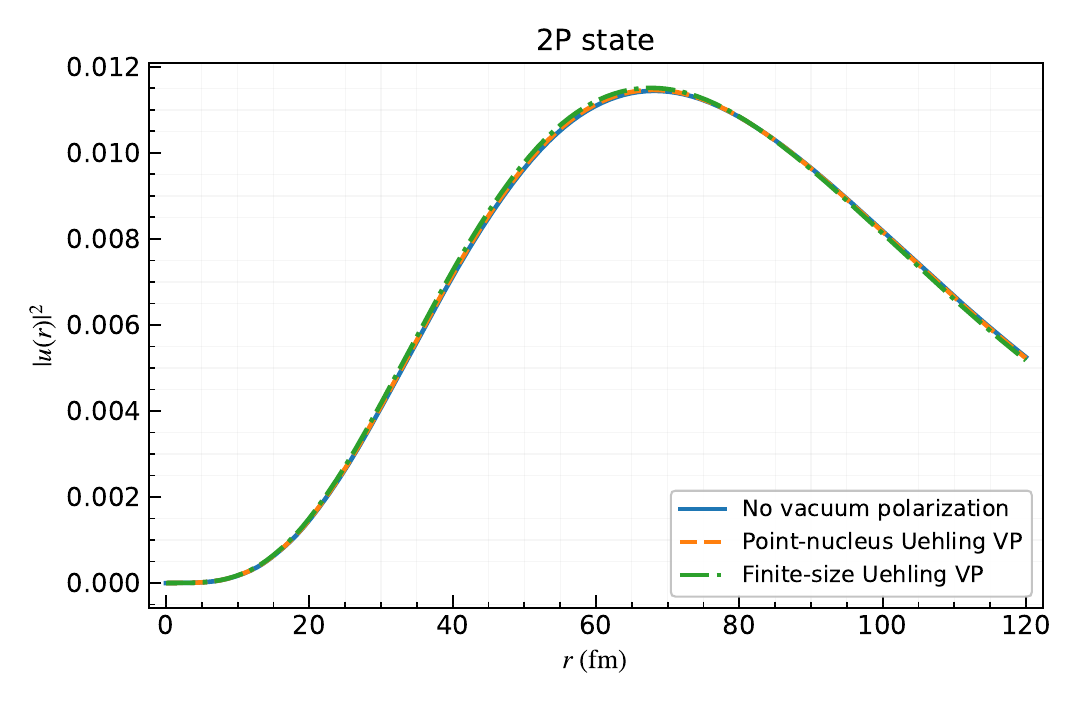}

\vspace{1mm}
{\small (b) $2P_{1/2}$ state}
\end{minipage}
\caption{%
Radial probability densities for the $2S_{1/2}$ and $2P_{1/2}$ states of
muonic $^{31}\mathrm{P}$ under three QED treatments: no vacuum polarization,
point-nucleus Uehling potential, and finite-size Uehling potential. Even for
$^{31}\mathrm{P}$, the heaviest system considered in the present validation
test, the three curves are nearly indistinguishable, showing that the
finite-size Uehling correction does not affect the qualitative conclusions of
the survey.}
\label{fig:wavefunction_comparison}
\end{figure}

To quantify the impact of this correction, we compare the radial probability
densities $|u(r)|^2 = r^2|\psi(r)|^2$ for the $2S_{1/2}$ and $2P_{1/2}$ states
of $\mu^{31}$P ($Z=15$, $R_{\rm ch}=3.19$~fm~\cite{Angeli2013}) under three
levels of approximation: (i) no vacuum polarization; (ii) point-nucleus Uehling;
and (iii) finite-size Uehling from Eq.~\eqref{eq:V_Ueh_radial_convolution}. As
shown in Fig.~\ref{fig:wavefunction_comparison}, the three curves are numerically
indistinguishable across the full radial range, including the region
$r \lesssim m_X^{-1} \approx 12$~fm where the $X17$ Yukawa potential is most
concentrated. The finite-size Uehling correction is known to be negligible for
light nuclei and to grow only for $Z \gtrsim 20$~\cite{Borie2012}; $^{31}$P at
$Z=15$ therefore represents the most demanding case in our survey.  We conclude
that the point-nucleus approximation is unlikely to alter the ranking of optimal
targets or the qualitative conclusions of the survey for any nucleus in our range.
Similarly, folding the short-range $X17$ Yukawa potential with the nuclear density
distribution $\rho_{\rm ch}$ would introduce corrections for the heaviest nuclei
($Z \sim 15$); we leave the evaluation of this folded Yukawa potential to future
precision studies, as it falls within the current margin of nuclear-structure
uncertainties.

\paragraph{Higher-order QED effects.}
Beyond the leading Uehling term, the Lamb shift in muonic atoms receives
contributions from the K\"{a}ll\'{e}n--Sabry two-loop vacuum polarization, the
Wichmann--Kroll self-energy correction, and hadronic vacuum polarization.  These
are collectively known to contribute at the level of $\sim$\,0.01--0.1\% of the
total Lamb shift in light muonic
atoms~\cite{Borie2012,Eides2007}.  Because these corrections arise from loop
insertions on the photon propagator, they modify the overall energy scale without
introducing length scales comparable to $m_X^{-1} \approx 12$~fm.  Their effect
on the short-range wave-function amplitude---which governs the $X17$-induced
hyperfine shifts---is therefore expected to remain subleading for the purpose of
target identification, and is unlikely to alter the ranking of optimal candidates
relative to the nuclear-spin uncertainty discussed below.

\paragraph{Nuclear-structure uncertainties.}
The dominant theoretical uncertainty in the $X17$-induced hyperfine shifts derives
from the Schmidt-model treatment of nuclear spin content, rather than from QED
corrections to the wave function.  Configuration mixing and core polarization can
shift the effective spin fractions $\Delta_{p,n}$ by amounts that are
nucleus-dependent and not uniformly
constrained~\cite{Stone2005,BohrMottelson1998,NeugartNeyens2017}.  Since our
primary goal is to identify the most sensitive experimental targets rather than to
make definitive quantitative predictions, this limitation does not affect the
principal conclusions of the survey.

\section{Results and Experimental Prospects}
\label{sec:results}

\subsection{Benchmark validation of the GEM framework}
\label{sec:bench}

Before presenting the $X17$-induced signals, we benchmark the GEM framework in
two complementary ways. First, we compare the electromagnetic baseline energies
with published high-precision QED results. Second, we compare the dimensionless
$X17$-induced hyperfine ratio $E_X/E_F$ with the analytic estimates of
Jentschura~\cite{Jentschura2020PRA} for $\mu$p and $\mu$d. The results are
summarized in Table~\ref{tab:benchmark}.

\begin{table}[H]
\centering
\caption{%
GEM results benchmarked against published calculations.
\textit{Upper block}: electromagnetic level energies vs.\ high-precision
QED; reference values from Faustov \& Martynenko~\cite{Martynenko2004JETP}
($1S$ HFS in $\mu$p), Faustov, Martynenko et al.~\cite{Krutov2014PRA} ($1S$ HFS
in $\mu$d), Martynenko~\cite{Martynenko2008JETP} ($1S$ HFS in
$\mu^3\mathrm{He}^+$), Borie~\cite{Borie2012} (Lamb shift in $\mu$p, total with
$R_p=0.875$~fm), Krutov \& Martynenko~\cite{Krutov2011PRA} (Lamb shift in $\mu$d).
\textit{Lower block}: dimensionless $X17$-to-Fermi ratio $E_X/E_F$ for the
$1S_{1/2}$ HFS vs.\ the analytic estimates of
Jentschura~\cite{Jentschura2020PRA} [Eq.~\eqref{eq:jentschura_ratio}],
evaluated with the coupling parameters adopted in this work.
EM deviations: $\delta = |E_{\rm GEM} - E_{\rm ref}|/|E_{\rm ref}|\times 100\%$.}
\label{tab:benchmark}
\setlength{\tabcolsep}{5pt}
\renewcommand{\arraystretch}{1.35}
\begin{tabular}{@{}llcccc@{}}
\toprule
Observable & System
& $E_{\rm GEM}$
& $E_{\rm ref}$
& $\delta$ [\%]
& Reference \\
\midrule
\multicolumn{6}{@{}l}{\textit{Electromagnetic baseline (energies in meV)}}\\[2pt]
\multirow{3}{*}{$1S_{1/2}$ HFS}
  & $\mu$p        & $183.235$ & $182.638$ & $0.33$ & \cite{Martynenko2004JETP} \\
  & $\mu$d        & $49.644$  & $48.584$  & $2.18$ & \cite{Krutov2014PRA}      \\
  & $\mu^3\mathrm{He}^+$ & $1377.59$ & $1334.56$ & $3.22$ & \cite{Martynenko2008JETP} \\[2pt]
\multirow{2}{*}{Lamb shift $|E(2S_{1/2}){-}E(2P_{1/2})|$}
  & $\mu$p  & $198.677$ & $202.05^\star$ & $1.67$ & \cite{Borie2012}     \\
  & $\mu$d  & $198.837$ & $202.414$      & $1.77$ & \cite{Krutov2011PRA} \\[4pt]
\midrule
\multicolumn{6}{@{}l}{\textit{$X17$-to-Fermi ratio $E_X/E_F$ }}\\[2pt]
\multirow{2}{*}{Vector $X17$, $1S_{1/2}$}
  & $\mu$p  & $5.6\times10^{-7}$ & $3.3\times10^{-8}$  & $\ddagger$ & Eq.~\eqref{eq:jentschura_ratio}a \\
  & $\mu$d  & $3.8\times10^{-6}$ & $1.4\times10^{-6}$  & $\ddagger$ & Eq.~\eqref{eq:jentschura_ratio}a \\[2pt]
\multirow{2}{*}{Pseudoscalar $X17$, $1S_{1/2}$}
  & $\mu$p  & $4.7\times10^{-6}$ & $1.3\times10^{-6}$  & $\ddagger$ & Eq.~\eqref{eq:jentschura_ratio}b \\
  & $\mu$d  & $3.6\times10^{-6}$ & $4.5\times10^{-6}$  & $\ddagger$ & Eq.~\eqref{eq:jentschura_ratio}b \\
\bottomrule
\end{tabular}
\vspace{3pt}
\begin{flushleft}
{\footnotesize
$^\star$ Evaluated from Borie's parameterization
$\Delta E_{LS} = 206.060 - 5.2794\langle r^2\rangle + 0.0546\langle r^2\rangle^{3/2}$~meV
with $R_p = 0.875$~fm ($\langle r^2\rangle = 0.7656$~fm$^2$). \\
$^\ddagger$ Both the GEM numerical ratios and the reference values are evaluated
with the coupling parameters adopted in this work. The residual factor-of-2--10
discrepancies arise from the approximations in the analytic formula
(Eq.~(30) of Ref.~\cite{Jentschura2020PRA}), which retains only the leading
contact-term limit $\sim\delta^{(3)}(\mathbf{r})$, whereas GEM integrates the
full Yukawa radial profile $e^{-m_X r}/r$. This order-of-magnitude agreement
validates the correct implementation of the $X17$--nucleon interaction in the
GEM Hamiltonian. \\
$E_F = E_{\rm HFS}^{\rm GEM}/4$ (spin-$\tfrac{1}{2}$) or
$E_{\rm HFS}^{\rm GEM}/6$ (spin-$1$ deuteron). \\
}
\end{flushleft}
\end{table}

For the electromagnetic baseline, the agreement degrades gradually with
increasing nuclear complexity: the deviation in the $1S_{1/2}$ hyperfine
splitting grows from $0.33\%$ in $\mu$p to $2.18\%$ in $\mu$d and
$3.22\%$ in $\mu^3\mathrm{He}^+$. This trend is consistent with the increasing
importance of higher-order nuclear-structure effects, such as Zemach,
polarizability, and relativistic recoil corrections, which are included in the
reference calculations~\cite{Martynenko2004JETP,Krutov2014PRA,Martynenko2008JETP}
but not in the present Hamiltonian. For the Lamb shift, both $\mu$p and $\mu$d
show a systematic deficit of about $3.4$~meV ($\sim 1.7\%$), which can be
attributed mainly to missing higher-order vacuum-polarization and recoil
contributions.

The lower block of Table~\ref{tab:benchmark} tests the new-physics sector by
comparing the numerical ratio
$E_X/E_F \equiv \Delta E^{X17}/E_{\rm Fermi}$ with the leading-order analytic
estimates of Jentschura~\cite{Jentschura2020PRA},
\begin{subequations}
\label{eq:jentschura_ratio}
\begin{align}
\frac{E_{X,V}(nS_{1/2})}{E_F(nS_{1/2})} &\approx
  -\frac{2 h'_\mu h'_N}{g_N \pi}\,\frac{Z m_r}{m_X}\,,
\label{eq:jentschura_ratio_a} \\
\frac{E_{X,A}(nS_{1/2})}{E_F(nS_{1/2})} &\approx
   \frac{h_\mu h_N}{g_N \pi}\,\frac{Z m_r}{m_X}\,.
\label{eq:jentschura_ratio_b}
\end{align}
\end{subequations}
The agreement is at the order-of-magnitude level, as expected, since the
analytic expressions retain only the leading contact-term limit, whereas the GEM
calculation includes the full Yukawa radial profile. This comparison confirms
that the $X17$ interaction has been implemented correctly in the numerical
framework.

Overall, for the benchmark systems $\mu$p, $\mu$d, and $\mu^3\mathrm{He}^+$,
the GEM reproduces the electromagnetic observables at the $0.3$--$3\%$ level,
which is sufficient for the purpose of target identification. For heavier
systems in the survey, analogous high-precision benchmarks are not available,
but the omitted higher-order QED contributions are expected to remain too small
to alter the ranking of optimal candidates. The dominant theoretical
uncertainty therefore comes from the Schmidt-model treatment of nuclear spin
content rather than from the GEM baseline itself.

% ──────────────────────────────────────────────────────────────────
\subsection{Vector-induced $2S_{1/2}$--$2P_{1/2}$ Lamb shifts}
\label{sec:lamb}

A vector $X17$ mediator induces a spin-independent Yukawa potential through the
coherent nuclear coupling
\begin{equation}
V_X(r) \;=\; \frac{h'_\mu}{4\pi}\,(Z h'_p + N h'_n)\,\frac{e^{-m_X r}}{r}.
\label{eq:VX_lamb_sec}
\end{equation}
This interaction shifts the level centroids, with the largest effect occurring
for $S$ states whose wave functions probe the nuclear region most strongly. In
contrast to the hyperfine sector, where the relevant couplings are weighted by
the spin-carrying valence nucleon(s), the Lamb shift depends on the fully
coherent combination $Zh_p' + Nh_n'$. Because the protophobic benchmark implies
$h_n' \gg h_p'$, the vector-induced Lamb-shift signal increases strongly toward
neutron-rich and heavier nuclei.

We define
\begin{equation}
\Delta E_{X17}^{(V)} \equiv
E_{X17}^{(V)}(2S_{1/2})-E_{X17}^{(V)}(2P_{1/2})
\label{eq:DeltaEX17V}
\end{equation}
and use its absolute value as the experimentally relevant quantity. The full
results for all stable nuclei with $Z\le 15$ are listed in
Appendix~\ref{app:lamb-vector} and shown in Fig.~\ref{fig:lamb_shift}.

A clear systematic pattern emerges across the nuclear chart. In the lightest
systems, the signal increases rapidly from hydrogenic to helium systems:
$|\Delta E_{X17}^{(V)}| = 2.05\times10^{-3}$~meV for $\mu p$,
$3.20\times10^{-2}$~meV for $\mu d$,
$0.245$~meV for $\mu^3\mathrm{He}^+$, and
$0.468$~meV for $\mu^4\mathrm{He}^+$. This growth reflects both the increasing
short-distance wave-function amplitude and the coherent enhancement from the
nuclear coupling.

Unlike hyperfine observables, the vector Lamb shift is not restricted to nuclei
with nonzero spin. As a result, even--even nuclei become fully competitive and
often dominant, although the strongest case in the present survey is an odd-$Z$
system. Large shifts are found already in
$\mu^{12}\mathrm{C}$ ($21.2$~meV),
$\mu^{13}\mathrm{C}$ ($24.5$~meV),
$\mu^{16}\mathrm{O}$ ($50.4$~meV),
$\mu^{18}\mathrm{O}$ ($62.1$~meV),
$\mu^{20}\mathrm{Ne}$ ($94.2$~meV),
$\mu^{24}\mathrm{Mg}$ ($150$~meV),
$\mu^{26}\mathrm{Mg}$ ($174$~meV),
$\mu^{28}\mathrm{Si}$ ($174$~meV), and
$\mu^{30}\mathrm{Si}$ ($197$~meV). The target hierarchy is therefore governed
primarily by coherent charge enhancement and radial overlap, rather than by
nuclear spin structure.

The $p$-shell nuclei form an intermediate region between the tiny hydrogenic
signals and the much larger $sd$-shell shifts. In particular,
$\mu^9\mathrm{Be}$, $\mu^{10}\mathrm{B}$, $\mu^{11}\mathrm{B}$, and
$\mu^{13}\mathrm{C}$ yield
$6.92$, $11.84$, $14.06$, and $24.5$~meV, respectively, while
$\mu^6\mathrm{Li}$ and $\mu^7\mathrm{Li}$ give $2.06$ and $2.71$~meV.

The strongest vector-induced Lamb-shift signal in the survey is obtained for
$\mu^{31}\mathrm{P}$, with
$|\Delta E_{X17}^{(V)}| = 264$~meV. The next strongest case is
$\mu^{29}\mathrm{Si}$ ($237$~meV), followed by
$\mu^{30}\mathrm{Si}$ ($197$~meV),
$\mu^{27}\mathrm{Al}$ ($195.5$~meV),
$\mu^{28}\mathrm{Si}$ ($174$~meV),
$\mu^{26}\mathrm{Mg}$ ($174$~meV),
$\mu^{25}\mathrm{Mg}$ ($161.7$~meV), and
$\mu^{24}\mathrm{Mg}$ ($150$~meV), with
$\mu^{21}\mathrm{Ne}$ still reaching $102.9$~meV. Although the ratio
$|\Delta E_{X17}^{(V)}|/|\Delta E_{L}^{(\mathrm{EM})}|$ decreases with
increasing $Z$, because the ordinary Lamb shift grows even faster, the absolute
$X17$ signal continues to rise over much of the survey range.

By contrast, a pseudoscalar $X17$ does not shift the spin-averaged level
centroids at leading order, since its interaction is purely spin dependent and
therefore averages to zero in the unpolarized $2S$ and $2P$ centroids. The Lamb
shift is therefore a particularly clean probe of the vector hypothesis.

\begin{figure}[t]
\centering
\includegraphics[width=\textwidth]{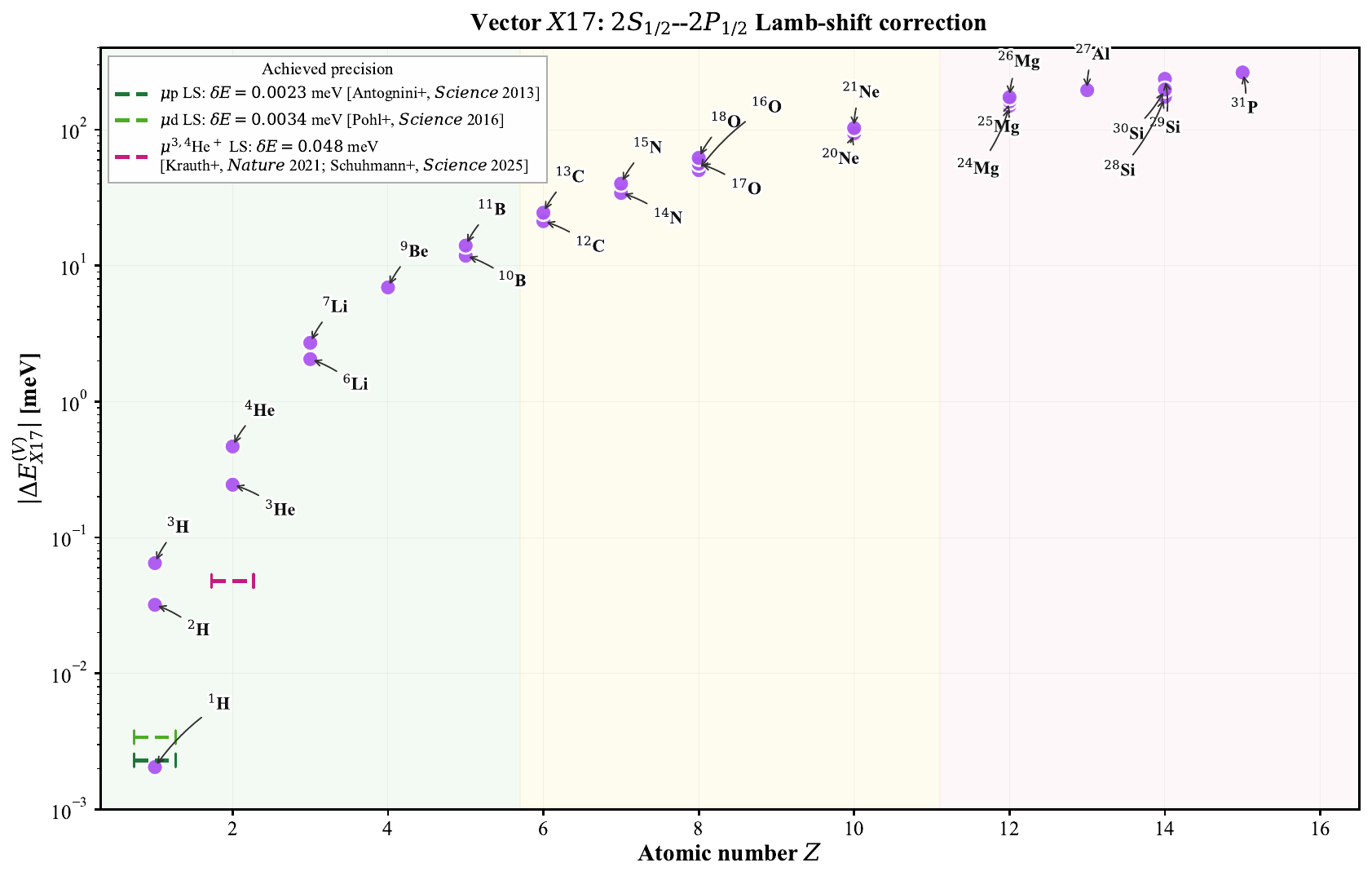}
\caption{%
Vector-$X17$ contribution to the $2S_{1/2}$--$2P_{1/2}$ Lamb shift for stable
muonic atoms with $Z\le 15$. The signal grows systematically toward heavier
nuclei and reaches its largest value in $\mu^{31}\mathrm{P}$ ($264$~meV).}
\label{fig:lamb_shift}
\end{figure}

\subsection{\texorpdfstring{$1S_{1/2}$}{1S1/2} hyperfine signatures}
\label{sec:hfs}

\begin{figure}[t]
\centering
\includegraphics[width=\textwidth]{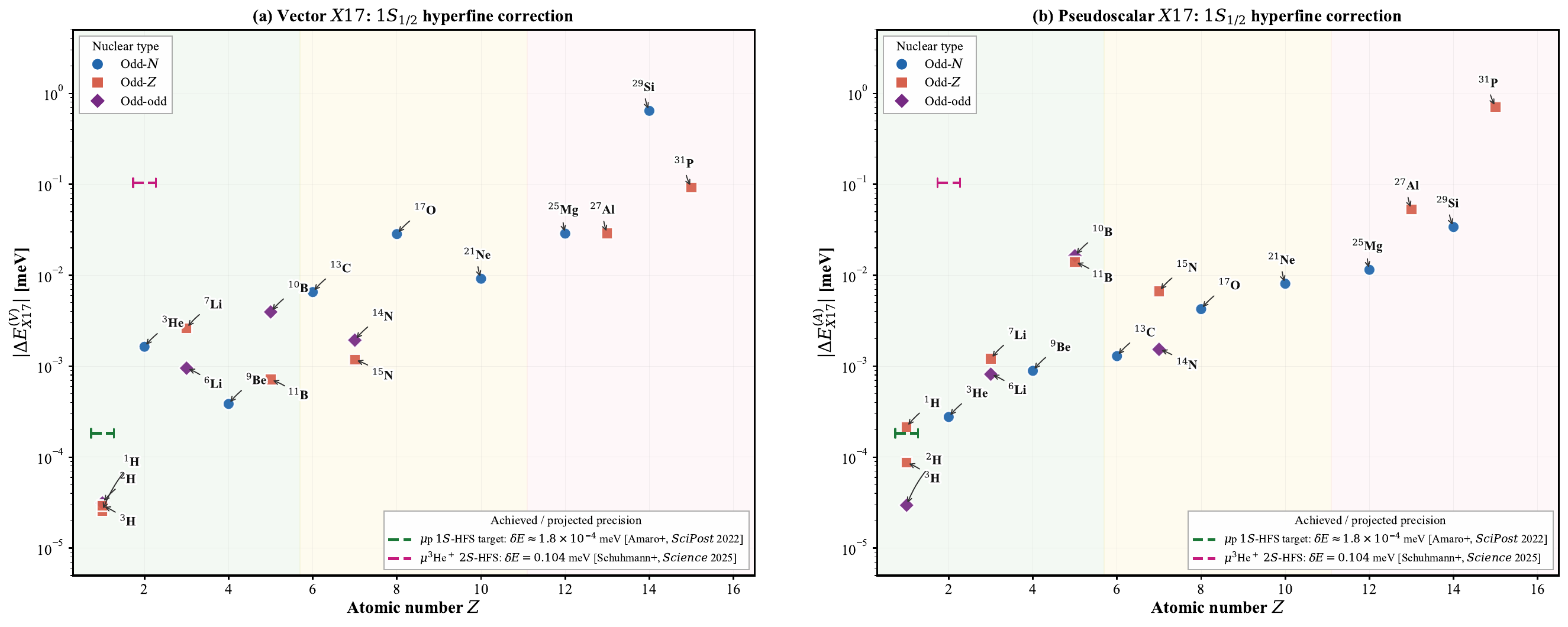}
\caption{%
$X17$-induced $1S_{1/2}$ hyperfine corrections for stable muonic atoms with
$Z\le 15$: (a) vector mediator and (b) pseudoscalar mediator. The vector case
preferentially enhances odd-$N$ nuclei, while the pseudoscalar case favors
odd-$Z$ nuclei. The largest signals are found in $\mu^{29}\mathrm{Si}$
($0.643$~meV) and $\mu^{31}\mathrm{P}$ ($0.710$~meV), respectively.}
\label{fig:hfs_1s}
\end{figure}

The $1S_{1/2}$ hyperfine splitting probes the spin structure of both the muon
and the nucleus, and therefore accesses coupling combinations that are distinct
from those entering the spin-independent Lamb shift. The full numerical results
are listed in Appendices~\ref{app:hfs-vector} and~\ref{app:hfs-ps}, and the
main trends are summarized in Fig.~\ref{fig:hfs_1s}.

For $1S$ states, the dominant $X17$ contribution arises from the short-range
contact term and therefore scales with the wave-function density at the origin,
\begin{equation}
|\psi_{1S}(0)|^2 \propto Z^3 m_r^3
\label{eq:psi1S_scaling}
\end{equation}
The overall size of the signal is then controlled by the effective nucleon-side
couplings, which depend on the nuclear spin fractions $\Delta_p$ and $\Delta_n$.
Unlike the vector Lamb shift, the hyperfine sector is therefore intrinsically
sensitive to nuclear spin structure and restricted to non-even-even nuclei.

For a vector mediator, the relevant effective coupling is
\[
h_N^{\prime(\mathrm{eff})}=h_p^\prime\Delta_p+h_n^\prime\Delta_n.
\]
Because $h_n^\prime \gg h_p^\prime$ in the protophobic benchmark adopted here,
the hierarchy is driven mainly by neutron spin content, and odd-$N$ nuclei are
strongly favored. The largest $1S_{1/2}$ hyperfine signal in the survey is found
in $\mu^{29}\mathrm{Si}$, where
$|\Delta E_{X17}^{(V)}| = 0.643$~meV. This case combines a relatively large
short-distance wave-function amplitude with maximal neutron-spin sensitivity
($\Delta_n=1$, $\Delta_p=0$). Other notable vector cases include
$\mu^{31}\mathrm{P}$, $\mu^{27}\mathrm{Al}$, $\mu^{25}\mathrm{Mg}$, and
$\mu^{17}\mathrm{O}$, although all remain well below the leading
$\mu^{29}\mathrm{Si}$ signal.

For a pseudoscalar mediator, the effective coupling becomes
\[
h_N^{(\mathrm{eff})}=h_p\Delta_p+h_n\Delta_n,
\]
and the hierarchy is reversed because the proton-side coupling dominates,
$|h_p| \gg |h_n|$. The pseudoscalar $1S_{1/2}$ hyperfine channel therefore
selects odd-$Z$ nuclei. The strongest signal is obtained in
$\mu^{31}\mathrm{P}$, with
$|\Delta E_{X17}^{(A)}| = 0.710$~meV, followed at a much lower level by
$\mu^{27}\mathrm{Al}$ at $0.053$~meV. In $\mu^{31}\mathrm{P}$, the valence
proton occupies the $2s_{1/2}$ orbital, giving $\Delta_p=1$ and $\Delta_n=0$,
which maximizes the pseudoscalar response.

The complementarity between the two hypotheses is therefore especially clear in
the $1S_{1/2}$ hyperfine sector: odd-$N$ nuclei dominate in the vector case,
whereas odd-$Z$ nuclei dominate in the pseudoscalar case. This provides a
natural discriminator between mediator scenarios.

\subsection{\texorpdfstring{$2P_{1/2}$}{2P1/2} hyperfine signatures}
\label{sec:hfs2p}

The $2P_{1/2}$ hyperfine corrections, shown in Fig.~\ref{fig:hfs_2p} and
tabulated in Appendices~\ref{app:hfs-vector:2p} and~\ref{app:hfs-ps:2p}, follow
the same qualitative selectivity as their $1S_{1/2}$ counterparts, but are
much smaller throughout the survey.

This suppression has two main origins. First, the $P$-state wave function
vanishes at the origin, so the short-range contact contribution that dominates
the $1S_{1/2}$ hyperfine splitting is strongly reduced. Since the Yukawa
interaction probes mainly the region
$r \lesssim m_X^{-1} \approx 12$~fm, the resulting $2P_{1/2}$ shifts are
suppressed by several orders of magnitude relative to the corresponding
$1S_{1/2}$ values. For the lightest systems, the vector-induced
$2P_{1/2}$ correction is already of order $10^{-9}$~meV.

Second, the $2P_{1/2}$ hyperfine operator contains tensor and spin--orbit terms
in addition to the reduced contact contribution, and these terms partially
cancel in many nuclei. An especially striking example is
$^{15}\mathrm{N}$, for which the vector-induced correction is only
$|\Delta E_{X17}^{(V)}(2P_{1/2})| \simeq 5.9\times10^{-10}$~meV because of an
almost complete cancellation among the relevant sublevel contributions.

Despite the strong suppression, the overall hierarchy remains consistent with
the $1S_{1/2}$ hyperfine sector. In the vector case, odd-$N$ nuclei still
dominate, with $\mu^{29}\mathrm{Si}$ giving the largest
$2P_{1/2}$ signal,
$|\Delta E_{X17}^{(V)}| = 8.70\times10^{-4}$~meV. In the pseudoscalar case,
odd-$Z$ nuclei remain favored, with $\mu^{31}\mathrm{P}$ reaching
$|\Delta E_{X17}^{(A)}| = 5.39\times10^{-4}$~meV. The persistence of the same
nuclear selectivity in both the $1S_{1/2}$ and $2P_{1/2}$ channels provides an
internal consistency check on the coupling pattern.

In absolute size, however, all $2P_{1/2}$ hyperfine shifts remain far below the
$1S_{1/2}$ signals. We therefore include them mainly for completeness and for
future theoretical cross-checks; their relevance for experimental target
selection is limited.

\begin{figure}[t]
\centering
\includegraphics[width=\textwidth]{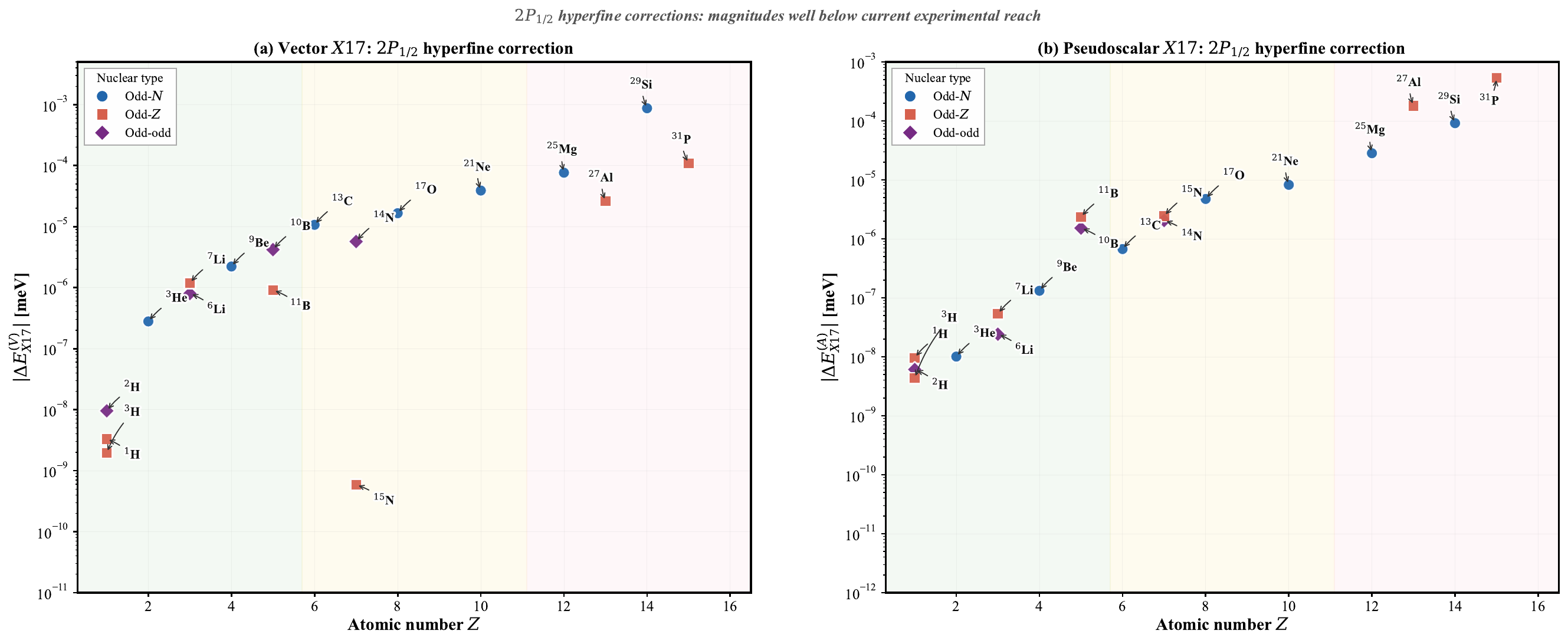}
\caption{%
$X17$-induced $2P_{1/2}$ hyperfine corrections for stable muonic atoms with
$Z\le 15$: (a) vector mediator and (b) pseudoscalar mediator. Compared with
the corresponding $1S_{1/2}$ channel, the corrections are strongly suppressed.
The largest signals are obtained in $\mu^{29}\mathrm{Si}$ for the vector case
and $\mu^{31}\mathrm{P}$ for the pseudoscalar case.}
\label{fig:hfs_2p}
\end{figure}

\subsection{Identification of optimal experimental targets}
\label{sec:targets}

Table~\ref{tab:targets_summary} summarizes the most relevant candidate systems
for the three experimentally meaningful observable--hypothesis combinations
considered here: the vector-$X17$ Lamb shift, the vector-$X17$
$1S_{1/2}$ hyperfine splitting, and the pseudoscalar-$X17$
$1S_{1/2}$ hyperfine splitting. For each channel, we list both the leading
theoretical candidates ranked by signal size and the light systems for which
precision benchmarks already exist. The $2P_{1/2}$ hyperfine channel is not
included, since its predicted signals remain far below current or foreseeable
experimental sensitivity.

Among systems with demonstrated precision spectroscopy, the most promising
near-term probes of the vector-induced Lamb shift are $\mu d$,
$\mu^3\mathrm{He}^+$, and $\mu^4\mathrm{He}^+$. Their predicted signals reach
$|\Delta E_{X17}^{(V)}| = 3.20\times10^{-2}$~meV, $0.245$~meV, and
$0.468$~meV, corresponding to
$\mathcal R \simeq 9.4$, $5.1$, and $9.8$, respectively, relative to existing
or demonstrated precision scales. These systems therefore offer the most direct
near-term experimental access to the vector Lamb-shift channel.

Beyond the light benchmark systems, the largest absolute vector Lamb-shift
signals are found in the upper part of the present survey range. The strongest
case is $\mu^{31}\mathrm{P}$, with
$|\Delta E_{X17}^{(V)}| = 264$~meV, followed by
$\mu^{29}\mathrm{Si}$, $\mu^{30}\mathrm{Si}$, $\mu^{27}\mathrm{Al}$,
$\mu^{28}\mathrm{Si}$, $\mu^{26}\mathrm{Mg}$, and $\mu^{25}\mathrm{Mg}$.
Although no meV-level spectroscopy benchmark yet exists for these nuclei, their
large predicted signals make them natural targets for future dedicated
programs. Between the light benchmark systems and these largest signals lies an
intermediate group---notably
$\mu^9\mathrm{Be}$, $\mu^{10}\mathrm{B}$, $\mu^{11}\mathrm{B}$, and
$\mu^{13}\mathrm{C}$---which may serve as useful stepping stones for future
experimental development.

For hyperfine observables, the leading long-term targets remain
$\mu^{29}\mathrm{Si}$ for the vector scenario and
$\mu^{31}\mathrm{P}$ for the pseudoscalar scenario. The corresponding
$1S_{1/2}$ signals are
$0.643$~meV and $0.710$~meV, respectively, and are the largest hyperfine
effects found in the survey. More generally, the target pattern is strongly
complementary: odd-$N$ nuclei are favored in the vector case, whereas odd-$Z$
nuclei dominate in the pseudoscalar case.

Muonic hydrogen remains an important benchmark. In the vector Lamb-shift
channel, the predicted signal in $\mu p$,
$|\Delta E_{X17}^{(V)}| = 2.05\times10^{-3}$~meV, lies slightly below the
achieved CREMA precision, corresponding to $\mathcal R=0.89$. By contrast, in
the pseudoscalar $1S_{1/2}$ hyperfine channel, the $\mu p$ signal reaches
$|\Delta E_{X17}^{(A)}| = 2.15\times10^{-4}$~meV, giving
$\mathcal R \simeq 1.2$ at the target precision of the ongoing PSI and RAL
programs. These light systems therefore remain essential anchors for
interpreting the heavier candidates.

Taken together, the present survey suggests a clear experimental hierarchy. For
the vector Lamb-shift channel, the most immediate opportunities are provided by
$\mu d$ and the helium ions, where the predicted signals already exceed
demonstrated precision benchmarks. For future dedicated programs, the largest
absolute gains lie in the heaviest systems considered here, especially
$\mu^{31}\mathrm{P}$ and the silicon, aluminum, and magnesium isotopes. In the
hyperfine sector, the complementary pair $\mu^{29}\mathrm{Si}$ (vector) and
$\mu^{31}\mathrm{P}$ (pseudoscalar) stands out as the most informative
combination for probing the spin-parity structure of a possible $X17$
mediator.

\begin{table}[H]
\centering
\caption{Representative experimental targets and validation benchmarks for the
$X17$ sensitivity survey in muonic atoms. For each observable--hypothesis
combination, systems are listed in descending order of predicted signal size.
The signal-to-precision ratio $\mathcal{R}\equiv |\Delta E_{X17}|/\delta E$
indicates whether a decisive test is feasible at the benchmark couplings adopted
in this work. The $2P_{1/2}$ HFS channel is not included because its predicted
signals remain far below current or foreseeable experimental sensitivity
(see Sec.~\ref{sec:hfs2p}).}
\label{tab:targets_summary}
\setlength{\tabcolsep}{5pt}
\renewcommand{\arraystretch}{1.30}
\begin{tabular}{@{}llccc l@{}}
\toprule
Observable & Nuclide & $|\Delta E_{X17}|$ [meV]
           & $\delta E$ [meV] & $\mathcal{R}$
           & Experimental context \\
\midrule

\multicolumn{6}{l}{\textit{Vector $X17$: $2S_{1/2}$--$2P_{1/2}$ Lamb shift}} \\[3pt]
& $\mu^{31}$P                 & $264$                 & ---                  & ---   & next-gen, proposed \\
& $\mu^{29}$Si                & $237$                 & ---                  & ---   & next-gen, proposed \\
& $\mu^{30}$Si                & $197$                 & ---                  & ---   & next-gen, proposed \\
& $\mu^{27}$Al                & $196$                 & ---                  & ---   & next-gen, proposed \\
& $\mu^{28}$Si                & $174$                 & ---                  & ---   & next-gen, proposed \\
& $\mu^{26}$Mg                & $174$                 & ---                  & ---   & next-gen, proposed \\
& $\mu^{4}\mathrm{He}^+$      & $0.468$               & $0.048$              & $9.75$ & \cite{Krauth2021Nature} \\
& $\mu^{3}\mathrm{He}^+$      & $0.245$               & $0.048$              & $5.10$ & \cite{Schuhmann2025Science} \\
& $\mu d$                     & $3.20\times10^{-2}$   & $3.4\times10^{-3}$   & $9.42$ & \cite{Pohl2016Science} \\
& $\mu p$                     & $2.05\times10^{-3}$   & $2.3\times10^{-3}$   & $0.89$ & \cite{Antognini2013Science} \\
\midrule

\multicolumn{6}{l}{\textit{Vector $X17$: $1S_{1/2}$ hyperfine splitting}} \\[3pt]
& $\mu^{29}$Si                & $0.643$               & ---                  & ---   & next-gen pulsed muon~\cite{Amaro2023SciPostHFS} \\
& $\mu^{31}$P                 & $0.092$               & ---                  & ---   & next-gen pulsed muon~\cite{Amaro2023SciPostHFS} \\
& $\mu^{27}$Al                & $0.029$               & ---                  & ---   & muonic X-ray spectroscopy \\
& $\mu^{25}$Mg                & $0.029$               & ---                  & ---   & muonic X-ray spectroscopy \\
& $\mu^{17}$O                 & $0.028$               & ---                  & ---   & muonic X-ray spectroscopy \\
& $\mu^{3}\mathrm{He}^+$      & $1.64\times10^{-3}$   & $0.104$              & $0.016$ & \cite{Schuhmann2025Science} \\
& $\mu p$                     & $2.57\times10^{-5}$   & $1.8\times10^{-4}$   & $0.14$ & \cite{Amaro2023SciPostHFS} \\
\midrule

\multicolumn{6}{l}{\textit{Pseudoscalar $X17$: $1S_{1/2}$ hyperfine splitting}} \\[3pt]
& $\mu^{31}$P                 & $0.710$               & ---                  & ---   & next-gen pulsed muon~\cite{Amaro2023SciPostHFS} \\
& $\mu^{27}$Al                & $0.053$               & ---                  & ---   & muonic X-ray spectroscopy \\
& $\mu^{10}$B                 & $0.016$               & ---                  & ---   & proposed, PSI \\
& $\mu^{11}$B                 & $0.014$               & ---                  & ---   & proposed, PSI \\
& $\mu^{15}$N                 & $0.007$               & ---                  & ---   & muonic X-ray spectroscopy \\
& $\mu^{3}\mathrm{He}^+$      & $2.77\times10^{-4}$   & $0.104$              & $0.003$ & \cite{Schuhmann2025Science} \\
& $\mu p$                     & $2.15\times10^{-4}$   & $1.8\times10^{-4}$   & $1.2$ & \cite{Amaro2023SciPostHFS} \\
\bottomrule
\end{tabular}
\end{table}

% ── 结论 ~~(˘ᵕ˘)✧~~
\section{Conclusions}
\label{sec:conclusion}

We have developed a systematic muonic-atom spectroscopy framework for probing
short-range MeV-scale interactions, and applied it to a nucleus-by-nucleus
survey of stable systems with $Z\le 15$. The main conceptual advance is that
spin-independent Lamb shifts and spin-dependent hyperfine splittings are treated
on different and physically appropriate nuclear footings within the same atomic
framework: the vector-induced Lamb shift is governed by the coherent
muon--nucleus coupling $Zh_p'+Nh_n'$, whereas the hyperfine sector is built
isotope by isotope from spin-fraction-weighted effective couplings. This
observable-dependent construction makes it possible to expose, in a controlled
way, the distinct nuclear systematics of vector and pseudoscalar mediators, and
thereby turns comparative muonic spectroscopy from a collection of isolated case
studies into a realistic target-selection strategy for new-physics searches.

A second important feature of the present work is that the benchmark muon-side
couplings are chosen in accordance with the latest muon $g\!-\!2$ landscape.
Rather than adopting benchmark scales motivated by the pre-2025 picture, we
adopt benchmark couplings defined with reference to the final Fermilab
measurement together with the 2025 Theory Initiative update, for which the
experimental--theoretical difference is substantially reduced. The resulting benchmarks should therefore not be viewed
as preferred fits to a persistent anomaly, but as conservative and
phenomenologically motivated reference scales for assessing the discovery reach
of precision muonic spectroscopy.

Within this framework, several robust conclusions emerge. In the spin-independent
channel, the vector-induced $2S_{1/2}$–$2P_{1/2}$ Lamb-shift signal exhibits
strong coherent enhancement and grows rapidly toward heavier nuclei. Among
systems with established precision benchmarks, $\mu d$, $\mu^3\mathrm{He}^+$,
and $\mu^4\mathrm{He}^+$ already stand out as the most promising near-term
probes, with signal-to-precision ratios $\mathcal{R}\simeq 9.4$, $5.1$, and
$9.8$, respectively. For future dedicated programs, the largest absolute
Lamb-shift signal is predicted in $\mu^{31}\mathrm{P}$, followed by several
nuclei in the P-Si-Al-Mg region, indicating that medium-mass muonic atoms are a
particularly favorable arena for short-range vector-force searches.

In the hyperfine sector, the survey reveals a clear and experimentally useful
complementarity between mediator hypotheses. Because the protophobic vector
benchmark is dominated by the neutron coupling, the vector scenario
preferentially selects odd-$N$ nuclei, with $\mu^{29}\mathrm{Si}$ emerging as
the leading $1S_{1/2}$ hyperfine target. By contrast, the pseudoscalar scenario
is driven primarily by proton spin content and therefore favors odd-$Z$ systems,
with $\mu^{31}\mathrm{P}$ yielding the largest predicted $1S_{1/2}$ hyperfine
signal in the present survey. This odd-$N$ versus odd-$Z$ selectivity is one of
the most important outcomes of the present work, because it shows that
comparative spectroscopy across different nuclei can do more than test for the
presence of a signal: it can help diagnose the coupling pattern and
spin-parity structure of the underlying mediator.

The broader significance of the present study extends beyond the specific $X17$
interpretation of the ATOMKI anomalies. At a more general level, what we have
constructed is a spectroscopy strategy for generic short-range Yukawa-type
interactions in muonic atoms. Owing to the strong spatial compression of muonic
wave functions, these systems are exceptionally sensitive to new forces with
ranges of order a few to tens of femtometers, and the observable-dependent
framework introduced here can be adapted straightforwardly to a wider class of
vector, pseudoscalar, or other light-mediator scenarios. In this sense, the
present work broadens the role of muonic-atom spectroscopy from a tool for
testing selected anomaly-driven models to a more general precision platform for
new-physics searches at the interface of atomic, nuclear, and particle physics.

The main theoretical limitation of the present survey lies in the Schmidt-model
treatment of nuclear spin content in the hyperfine sector. While this level of
description is sufficient for establishing the global hierarchy of targets,
configuration mixing and core polarization can modify the effective spin
fractions for individual nuclei, especially in the medium-mass region.
Accordingly, the most important next theoretical step is to combine the present
atomic framework with improved nuclear-structure calculations for the leading
hyperfine candidates. On the experimental side, the results point to a natural
two-stage program: near-term tests in $\mu d$ and the helium sector, where
precision benchmarks already exist, followed by longer-term dedicated studies of
the most favorable heavier nuclei in the present survey, especially
$\mu^{31}\mathrm{P}$ and $\mu^{29}\mathrm{Si}$.

Overall, the present survey shows that precision muonic-atom spectroscopy can
provide not only competitive sensitivity to short-range fifth-force effects, but
also genuine diagnostic power on their coupling structure. This makes muonic
atoms a uniquely valuable laboratory for exploring light new interactions
beyond the Standard Model.
	\section*{Acknowledgments}
	This work has been supported by the Research Program of State Key Laboratory of Heavy Ion Science and Technology, Institute of Modern Physics, Chinese Academy of Sciences (Grant No. HIST2025CS08), and the National Key R$\&$D Program of China (Grant No. 2024YFE0109800 and 2024YFE0109802).
	\bibliographystyle{apsrev4-2}
	\bibliography{reference.bib}

@article{Krasznahorkay2016PRL,
	author       = {Krasznahorkay, A. J. and others},
	title        = {Observation of Anomalous Internal Pair Creation in $^8$Be: A Possible Indication of a Light, Neutral Boson},
	journal      = {Phys. Rev. Lett.},
	volume       = {116},
	pages        = {042501},
	year         = {2016},
	doi          = {10.1103/PhysRevLett.116.042501},
	eprint       = {1504.01527},
	archivePrefix= {arXiv}
}

@article{Be4_Krasz_2019,
	author       = {Krasznahorkay, A. J. and others},
	title        = {New evidence supporting the existence of the hypothetic X17 particle},
	journal      = {arXiv preprint},
	eprint       = {1910.10459},
	year         = {2019},
	archivePrefix= {arXiv},
	note         = {Anomalous $e^+e^-$ pairs in $^4$He transitions}
}

@article{Feng2016PRL,
	author  = {Feng, Jonathan L. and Fornal, Bartosz and Galon, Iftah and Gardner, Susan and Smolinsky, Jordan and Tait, Tim M. P. and Tanedo, Philip},
	title   = {Protophobic Fifth-Force Interpretation of the Observed Anomaly in $^8$Be Nuclear Transitions},
	journal = {Phys. Rev. Lett.},
	volume  = {117},
	pages   = {071803},
	year    = {2016},
	doi     = {10.1103/PhysRevLett.117.071803},
	eprint  = {1604.07411},
	archivePrefix= {arXiv}
}

@book{Eides2007,
	author    = {Eides, M. I. and Grotch, H. and Shelyuto, V. A.},
	title     = {Theory of Light Hydrogenlike Atoms},
	series    = {Springer Tracts in Modern Physics},
	volume    = {222},
	year      = {2007},
	doi       = {10.1007/3-540-45270-1}
}

@article{Hiyama2003,
	author  = {Hiyama, E. and Kino, Y. and Kamimura, M.},
	title   = {Gaussian Expansion Method for Few-Body Systems},
	journal = {Progress in Particle and Nuclear Physics},
	volume  = {51},
	pages   = {223--307},
	year    = {2003},
	doi     = {10.1016/S0146-6410(03)90015-9}
}

@book{KraneBook,
	author    = {Kenneth S. Krane},
	title     = {Introductory Nuclear Physics},
	publisher = {Wiley},
	address   = {New York},
	year      = {1987},
	isbn      = {978-0471805533}
}

@article{AME2020,
	author  = {Wang, Meng and Huang, W. J. and Kondev, F. G. and Audi, G. and Naimi, S.},
	title   = {The AME2020 atomic mass evaluation},
	journal = {Chin. Phys. C},
	volume  = {45},
	number  = {3},
	pages   = {030003},
	year    = {2021},
	doi     = {10.1088/1674-1137/abddaf}
}

@article{Leveille1978,
	author       = {J. P. Leveille},
	title        = {The Second-Order Weak Correction to $(g-2)$ of the Muon in Arbitrary Gauge Models},
	journal      = {Nucl. Phys. B},
	volume       = {137},
	pages        = {63--76},
	year         = {1978},
	doi          = {10.1016/0550-3213(78)90291-7}
}

@article{LindnerPlatscherQueiroz2018,
	author       = {M. Lindner and M. Platscher and F. S. Queiroz},
	title        = {A Call for New Physics: The Muon Anomalous Magnetic Moment and Lepton Flavor Violation},
	journal      = {Phys. Rept.},
	volume       = {731},
	pages        = {1--82},
	year         = {2018},
	doi          = {10.1016/j.physrep.2017.12.001},
	eprint       = {1610.06587},
	archivePrefix= {arXiv}
}

@article{Jentschura2020PRA,
	author  = {Ulrich D. Jentschura},
	title   = {Fifth Force and Hyperfine Splitting in Bound Systems},
	journal = {Phys. Rev. A},
	volume  = {101},
	pages   = {062510},
	year    = {2020},
	doi     = {10.1103/PhysRevA.101.062510},
	eprint  = {2003.07207},
	archivePrefix = {arXiv}
}

@article{NeugartNeyens2017,
	author  = {R. Neugart and G. Neyens},
	title   = {Nuclear Moments},
	journal = {Journal of Physics G: Nuclear and Particle Physics},
	year    = {2017}
}

@article{Stone2005,
	author  = {N. J. Stone},
	title   = {Table of Nuclear Magnetic Dipole and Electric Quadrupole Moments},
	journal = {Atomic Data and Nuclear Data Tables},
	volume  = {90},
	pages   = {75--176},
	year    = {2005},
	doi     = {10.1016/j.adt.2005.04.001}
}

@book{BohrMottelson1998,
	author    = {A. Bohr and B. R. Mottelson},
	title     = {Nuclear Structure, Vols. I \& II},
	publisher = {World Scientific},
	year      = {1998},
	note      = {Reprint edition}
}

@article{Mayer1949,
	author  = {Maria Goeppert Mayer},
	title   = {On Closed Shells in Nuclei},
	journal = {Physical Review},
	year    = {1949}
}

@article{Jensen1955,
	author  = {J. Hans D. Jensen},
	title   = {The Nuclear Shell Model},
	journal = {Reviews of Modern Physics},
	year    = {1955}
}

@misc{KelleyTUNL,
	author = {J. H. Kelley and TUNL collaborators},
	title  = {TUNL Nuclear Data Evaluations (ENSDF/TUNL)},
	howpublished = {\url{https://nucldata.tunl.duke.edu/}}
}

@article{Krauth2017Science,
	author  = {Krauth, J. J. and Diepold, M. and Franke, B. and others (CREMA Collaboration)},
	title   = {The deuteron radius puzzle is solved by muonic deuterium},
	journal = {Science},
	year    = {2017},
	volume  = {358},
	number  = {6358},
	pages   = {79--85},
	doi     = {10.1126/science.aah6677}
}

@article{Angeli2013,
	author  = {Angeli, I. and Marinova, K. P.},
	title   = {Table of experimental nuclear ground state charge radii: An update},
	journal = {Atomic Data and Nuclear Data Tables},
	year    = {2013},
	volume  = {99},
	number  = {1},
	pages   = {69--95},
	doi     = {10.1016/j.adt.2011.12.006}
}

@misc{NNDC_NuDat3,
	title   = {NuDat~3.0 Database},
	author  = {{National Nuclear Data Center (NNDC)}},
	howpublished = {\url{https://www.nndc.bnl.gov/nudat3/}}
}

@misc{KAERI_Chart,
	title   = {Chart of the Nuclides},
	author  = {{Korea Atomic Energy Research Institute (KAERI)}},
	howpublished = {\url{https://atom.kaeri.re.kr/ton/}}
}

@techreport{Stone2016_Qmom,
	author  = {N. J. Stone},
	title   = {Table of Nuclear Electric Quadrupole Moments},
	institution = {IAEA Nuclear Data Section},
	number  = {INDC(NDS)-0650},
	year    = {2016},
	url     = {https://www-nds.iaea.org/publications/indc/indc-nds-0650/}
}

@article{PDG2024,
	title        = {Review of Particle Physics},
	author       = {Navas, S. and others},
	collaboration= {Particle Data Group},
	journal      = {Physical Review D},
	year         = {2024},
	volume       = {110},
	eid          = {030001},
	pages        = {030001},
	doi          = {10.1103/PhysRevD.110.030001}
}

@article{Jentschura2006PRA,
	author  = {U. D. Jentschura and V. A. Yerokhin},
	title   = {Quantum electrodynamic corrections to the hyperfine structure of excited S states},
	journal = {Physical Review A},
	volume  = {73},
	pages   = {062503},
	year    = {2006},
	doi     = {10.1103/PhysRevA.73.062503}
}

@article{Kamimura1988,
	author  = {Masayasu Kamimura},
	title   = {Nonadiabatic Coupled-Rearrangement-Channel Approach to Muonic Molecules. I. Formulation},
	journal = {Physical Review A},
	year    = {1988},
	volume  = {38},
	number  = {2},
	pages   = {621--624},
	doi     = {10.1103/PhysRevA.38.621}
}

@book{SuzukiVarga1998,
	author    = {Yasuyuki Suzuki and K{\'a}lm{\'a}n Varga},
	title     = {Stochastic Variational Approach to Quantum-Mechanical Few-Body Problems},
	series    = {Lecture Notes in Physics Monographs},
	volume    = {54},
	publisher = {Springer},
	address   = {Berlin, Heidelberg},
	year      = {1998},
	doi       = {10.1007/3-540-49541-X}
}

@article{Uehling:1935,
	author  = {Uehling, E. A.},
	title   = {Polarization Effects in the Positron Theory},
	journal = {Physical Review},
	volume  = {48},
	pages   = {55--63},
	year    = {1935},
	doi     = {10.1103/PhysRev.48.55}
}

@article{MuonG2_2025,
  title={Measurement of the positive muon anomalous magnetic moment to 127 ppb},
  author={Muon g- 2 Collaboration and others},
  journal={Physical Review Letters},
  volume={135},
  number={10},
  pages={101802},
  year={2025},
  doi={10.1103/7clf-sm2v}
}

@article{TI_WP25,
  title={The anomalous magnetic moment of the muon in the Standard Model: an update},
  author={Aliberti, R and Aoyama, T and Balzani, E and Bashir, A and Benton, G and Bijnens, J and Biloshytskyi, V and Blum, T and Boito, D and Bruno, M and others},
  journal={arXiv preprint arXiv:2505.21476},
  year={2025}
}

@article{Fullerton1976,
  title={Accurate and efficient methods for the evaluation of vacuum-polarization potentials of order Z $\alpha$ and Z $\alpha$ 2},
  author={Fullerton, L Wayne and Rinker Jr, GA},
  journal={Physical Review A},
  volume={13},
  number={3},
  pages={1283},
  year={1976},
  publisher={APS},
 doi     = {10.1103/PhysRevA.13.1283}
}

@article{Borie1982,
  title={The energy levels of muonic atoms},
  author={Borie, E and Rinker, GA},
  journal={Reviews of Modern Physics},
  volume={54},
  number={1},
  pages   = {67--118},
  year={1982},
  publisher={APS},
  doi     = {10.1103/RevModPhys.54.67}
}

@article{Ellwanger2016JHEP,
  title={Possible explanation of the electron positron anomaly at 17 MeV in 8 Be transitions through a light pseudoscalar},
  author={Ellwanger, Ulrich and Moretti, Stefano},
  journal={Journal of High Energy Physics},
  volume={2016},
  number={11},
  pages={1--13},
  year={2016},
  publisher={Springer},
  doi={10.1007/JHEP11(2016)039}
}

@article{Kozaczuk2017PRD,
  title={Light axial vector bosons, nuclear transitions, and the Be 8 anomaly},
  author={Kozaczuk, Jonathan and Morrissey, David E and Stroberg, SR},
  journal={Physical Review D},
  volume={95},
  number={11},
  pages={115024},
  year={2017},
  publisher={APS},
  doi={10.1103/PhysRevD.95.115024}
}

@article{Krutov2011EPJD,
  author  = {Krutov, A. A. and Martynenko, A. P.},
  title   = {Hyperfine structure of the ground state muonic $^{3}$He atom},
  journal = {Eur. Phys. J. D},
  volume  = {62},
  pages   = {163--175},
  year    = {2011},
  doi     = {10.1140/epjd/e2011-10401-5}
}

@article{Antognini2013Science,
  title={Proton structure from the measurement of 2S-2P transition frequencies of muonic hydrogen},
  author={Antognini, Aldo and Nez, Fran{\c{c}}ois and Schuhmann, Karsten and Amaro, Fernando D and Biraben, Francois and Cardoso, Joao MR and Covita, Daniel S and Dax, Andreas and Dhawan, Satish and Diepold, Marc and others},
  journal={Science},
  volume={339},
  number={6118},
  pages={417--420},
  year={2013},
  publisher={American Association for the Advancement of Science},
  doi={10.1126/science.1230016}
}

@article{Pohl2016Science,
  title={Laser spectroscopy of muonic deuterium},
  author={Pohl, Randolf and Nez, Fran{\c{c}}ois and Fernandes, Luis MP and Amaro, Fernando D and Biraben, Fran{\c{c}}ois and Cardoso, Joao MR and Covita, Daniel S and Dax, Andreas and Dhawan, Satish and Diepold, Marc and others},
  journal={Science},
  volume={353},
  number={6300},
  pages={669--673},
  year={2016},
  publisher={American Association for the Advancement of Science},
  doi     = {10.1126/science.aaf2468},
}

@article{Krauth2021Nature,
  title={Measuring the $\alpha$-particle charge radius with muonic helium-4 ions},
  author={Krauth, Julian J and Schuhmann, Karsten and Ahmed, Marwan Abdou and Amaro, Fernando D and Amaro, Pedro and Biraben, Fran{\c{c}}ois and Chen, Tzu-Ling and Covita, Daniel S and Dax, Andreas J and Diepold, Marc and others},
  journal={Nature},
  volume={589},
  number={7843},
  pages={527--531},
  year={2021},
  publisher={Nature Publishing Group UK London},
    doi     = {10.1038/s41586-021-03183-1},
}

@article{Schuhmann2025Science,
  title={The helion charge radius from laser spectroscopy of muonic helium-3 ions},
  author={Schuhmann, Karsten and Fernandes, Luis MP and Nez, Fran{\c{c}}ois and Abdou Ahmed, Marwan and Amaro, Fernando D and Amaro, Pedro and Biraben, Fran{\c{c}}ois and Chen, Tzu-Ling and Covita, Daniel S and Dax, Andreas J and others},
  journal={Science},
  volume={388},
  number={6749},
  pages={854--858},
  year={2025},
  publisher={American Association for the Advancement of Science},
    doi     = {10.1126/science.adj2610}
}

@article{Amaro2023SciPostHFS,
  title={Diffusion of muonic hydrogen in hydrogen gas and the measurement of the 1$ s $ hyperfine splitting of muonic hydrogen},
  author={Nuber, Jonas and Adamczak, A and Abdou Ahmed, M and Affolter, Lukas and Amaro, FD and Amaro, Pedro and Antognini, A and Carvalho, P and Chang, Y-H and Chen, T-L and others},
  journal={SciPost Physics Core},
  volume={6},
  number={3},
  pages={057},
  year={2022},
    doi     = {10.21468/SciPostPhysCore.6.3.057},
}

@article{Martynenko2004JETP,
  author    = {Faustov, R. N. and Martynenko, A. P.},
  title     = {Muonic hydrogen ground state hyperfine splitting},
  journal   = {J. Exp. Theor. Phys.},
  volume    = {98},
  number    = {1},
  pages     = {39--48},
  year      = {2004},
  doi       = {10.1134/1.1648079},
}

@article{Krutov2014PRA,
  author    = {Faustov, R. N. and Martynenko, A. P. and Martynenko, G. A. and Sorokin, V. V.},
  title     = {Hyperfine structure of {S}-states in muonic deuterium},
  journal   = {Phys. Rev. A},
  volume    = {90},
  pages     = {012520},
  year      = {2014},
  doi       = {10.1103/PhysRevA.90.012520},
}

@article{Martynenko2008JETP,
  author    = {Martynenko, A. P.},
  title     = {Hyperfine structure of the {S} levels of the muonic helium ion},
  journal   = {J. Exp. Theor. Phys.},
  volume    = {106},
  number    = {4},
  pages     = {690--699},
  year      = {2008},
  doi       = {10.1134/S1063776108040079},
}

@article{Borie2012,
  author    = {Borie, E.},
  title     = {Lamb shift in light muonic atoms---Revisited},
  journal   = {Ann. Phys.},
  volume    = {327},
  pages     = {733--763},
  year      = {2012},
  doi       = {10.1016/j.aop.2011.11.017}
}

@article{Krutov2011PRA,
  author    = {Krutov, A. A. and Martynenko, A. P.},
  title     = {Lamb shift in the muonic deuterium atom},
  journal   = {Phys. Rev. A},
  volume    = {84},
  pages     = {052514},
  year      = {2011},
  doi       = {10.1103/PhysRevA.84.052514}
}

@article{Martynenko2008PAN,
  author    = {Martynenko, A. P.},
  title     = {Fine and hyperfine structure of {P}-wave levels in muonic hydrogen},
  journal   = {Phys. At. Nucl.},
  volume    = {71},
  number    = {1},
  pages     = {125--135},
  year      = {2008},
  doi       = {10.1134/S1063778808010146}
}
	
% ── 附录 ~~✿~~
\appendix

\section{Nuclear inputs and spin structure}
\label{app:inputs}
\renewcommand{\thetable}{A\arabic{table}}
\setcounter{table}{0}

\subsection{Data sources and conventions}
\label{app:inputs:sources}
	\begin{longtable}{@{\extracolsep{\fill}} ccccccccc}
		\caption{Ground-state $J^\pi$, nuclear masses $m_{\rm nuc}$ [u], rms charge radii $r_{\rm ch}$ [fm; third column], and single-particle spin-sum projections (neutrons/protons) for stable nuclei with $Z\le 15$. The $r_{\rm ch}$ values for $^{1}\mathrm{H}$, $^{2}\mathrm{H}$, and $^{3}\mathrm{He}$ follow PDG/primary determinations \cite{Krauth2017Science,Schuhmann2025Science}; all other radii are taken from Angeli \& Marinova \cite{Angeli2013}. For even--even nuclei with $I=0$, the valence-orbital assignment is indicated as closed shell, while the spin-projection entries are marked by em dashes because no spin-dependent hyperfine contribution arises in the present framework.} \\
		\hline
		Nuclide & $m_{\text{nuc}}$ [u] &$\sqrt{\langle r^{2}\rangle}$ (fm)& $(Z,N)$ & Type & Ground $J^\pi$ & Valence orbital & $\displaystyle\sum_i \mathbf{S}_n^i$ & $\displaystyle\sum_i \mathbf{S}_p^i$ \\
		\hline
		\endfirsthead
		\hline
		\endhead
		\hline
		\endfoot
		\hline
		\endlastfoot
		
		$^{1}\mathrm{H}$  & 1.00727647 & 0.8409 & (1,0) & Odd Z & $\tfrac{1}{2}^{+}$ & p:$1s_{1/2}$ & $0$ & $+\tfrac{1}{2}\,\hbar\,\hat{\mathbf J}$ \\[2pt]
		$^{2}\mathrm{H}$  & 2.01355321 & 2.12799 & (1,1) & Odd–Odd & $1^{+}$ & p:$1s_{1/2}$;n:$1s_{1/2}$ & $+\tfrac{1}{2}\,\hbar\,\hat{\mathbf J}$ & $+\tfrac{1}{2}\,\hbar\,\hat{\mathbf J}$ \\[2pt]
		$^{3}\mathrm{H}$  & 3.01550071 & 1.759 & (1,2) & Odd Z & $\tfrac{1}{2}^{+}$ & p:$1s_{1/2}$ & $0$ & $+\tfrac{1}{2}\,\hbar\,\hat{\mathbf J}$ \\[2pt]
		$^{3}\mathrm{He}$ & 3.01493224 & 1.97007 & (2,1) & Odd N & $\tfrac{1}{2}^{+}$ & n:$1s_{1/2}$ & $+\tfrac{1}{2}\,\hbar\,\hat{\mathbf J}$ & $0$ \\[2pt]
		$^{4}\mathrm{He}$  & 4.00150618 & 1.681 & (2,2) & Even–Even & $0^{+}$ & closed shell & — & — \\[2pt]
		$^{6}\mathrm{Li}$  & 6.01347736 & 2.589 & (3,3) & Odd–Odd & $1^{+}$ & p:$1p_{3/2}$;n:$1p_{3/2}$ & $+\tfrac{1}{2}\,\hbar\,\hat{\mathbf J}$ & $+\tfrac{1}{2}\,\hbar\,\hat{\mathbf J}$ \\[2pt]
		$^{7}\mathrm{Li}$  & 7.01435791 & 2.444& (3,4) & Odd Z & $\tfrac{3}{2}^{-}$ & p:$1p_{3/2}$ & $0$ & $+\tfrac{1}{2}\,\hbar\,\hat{\mathbf J}$ \\[2pt]
		$^{9}\mathrm{Be}$  & 9.00998917 & 2.519 & (4,5) & Odd N & $\tfrac{3}{2}^{-}$ & n:$1p_{3/2}$ & $+\tfrac{1}{2}\,\hbar\,\hat{\mathbf J}$ & $0$ \\[2pt]
		$^{10}\mathrm{B}$ & 10.01019478 & 2.427 & (5,5) & Odd–Odd & $3^{+}$ & p:$1p_{3/2}$; n:$1p_{3/2}$ & $+\tfrac{1}{2}\,\hbar\,\hat{\mathbf J}$ & $+\tfrac{1}{2}\,\hbar\,\hat{\mathbf J}$ \\[2pt]
		$^{11}\mathrm{B}$ & 11.00656319 & 2.406 & (5,6) & Odd Z & $\tfrac{3}{2}^{-}$ & p:$1p_{3/2}$ & $0$ & $+\tfrac{1}{2}\,\hbar\,\hat{\mathbf J}$ \\[2pt]
		$^{12}\mathrm{C}$ & 11.99670964 & 2.470 & (6,6) & Even–Even & $0^{+}$ & closed shell & — & — \\[2pt]
		$^{13}\mathrm{C}$ & 13.00006448 & 2.461 & (6,7) & Odd N & $\tfrac{1}{2}^{-}$ & n:$1p_{1/2}$ & $-\tfrac{1}{6}\,\hbar\,\hat{\mathbf J}$ & $0$ \\[2pt]
		$^{14}\mathrm{N}$ & 13.99923557 & 2.558 & (7,7) & Odd–Odd & $1^{+}$ & p:$1p_{1/2}$; n:$1p_{1/2}$ & $-\tfrac{1}{6}\,\hbar\,\hat{\mathbf J}$ & $-\tfrac{1}{6}\,\hbar\,\hat{\mathbf J}$ \\[2pt]
		$^{15}\mathrm{N}$ & 14.99627046 & 2.612 & (7,8) & Odd Z & $\tfrac{1}{2}^{-}$ & p:$1p_{1/2}$ & $0$ & $-\tfrac{1}{6}\,\hbar\,\hat{\mathbf J}$ \\[2pt]
		$^{16}\mathrm{O}$ & 15.99052821 & 2.699 & (8,8) & Even–Even & $0^{+}$ & closed shell & — & — \\[2pt]
		$^{17}\mathrm{O}$ & 16.99474535 & 2.73 & (8,9) & Odd N & $\tfrac{5}{2}^{+}$ & n:$1d_{5/2}$ & $+\tfrac{1}{2}\,\hbar\,\hat{\mathbf J}$ & $0$ \\[2pt]
		$^{18}\mathrm{O}$ & 17.99477321 & 2.775 & (8,10) & Even–Even & $0^{+}$ & closed shell & — & — \\[2pt]
		$^{20}\mathrm{Ne}$ & 19.98695818 & 3.006 & (10,10) & Even–Even & $0^{+}$ & closed shell & — & — \\[2pt]
		$^{21}\mathrm{Ne}$ & 20.98836469 & 3.00 & (10,11) & Odd N & $\tfrac{3}{2}^{+}$ & n:$1d_{3/2}$ & $-\tfrac{3}{10}\,\hbar\,\hat{\mathbf J}$ & $0$ \\[2pt]
		$^{24}\mathrm{Mg}$ & 23.97846462 & 3.057 & (12,12) & Even–Even & $0^{+}$ & closed shell & — & — \\[2pt]
		$^{25}\mathrm{Mg}$ & 24.97925990 & 3.06 & (12,13) & Odd N & $\tfrac{5}{2}^{+}$ & n:$1d_{5/2}$ & $+\tfrac{1}{2}\,\hbar\,\hat{\mathbf J}$ & $0$ \\[2pt]
		$^{26}\mathrm{Mg}$ & 25.97601589 & 3.034 & (12,14) & Even–Even & $0^{+}$ & closed shell & — & — \\[2pt]
		$^{27}\mathrm{Al}$ & 26.97441412 & 3.061& (13,14) & Odd Z & $\tfrac{5}{2}^{+}$ & p:$1d_{5/2}$ & $0$ & $+\tfrac{1}{2}\,\hbar\,\hat{\mathbf J}$ \\[2pt]
		$^{28}\mathrm{Si}$ & 27.96925492 & 3.122 & (14,14) & Even–Even & $0^{+}$ & closed shell & — & — \\[2pt]
		$^{29}\mathrm{Si}$ & 28.96882305 & 3.13 & (14,15) & Odd N & $\tfrac{1}{2}^{+}$ & n:$2s_{1/2}$ & $+\tfrac{1}{2}\,\hbar\,\hat{\mathbf J}$ & $0$ \\[2pt]
		$^{30}\mathrm{Si}$ & 29.96609852 & 3.134 & (14,16) & Even–Even & $0^{+}$ & closed shell & — & — \\[2pt]
		$^{31}\mathrm{P}$  & 30.96554333 & 3.190 & (15,16) & Odd Z & $\tfrac{1}{2}^{+}$ & p:$2s_{1/2}$ & $0$ & $+\tfrac{1}{2}\,\hbar\,\hat{\mathbf J}$ \\[2pt]
	\end{longtable}

% ── ★ App A.2: 新增 Zh_p' + Nh_n' 列用于 Lamb shift ~~(｡•̀ᴗ-)✧~~ ──
\subsection{Spin content and effective couplings}
\label{app:inputs:spin}

\begin{longtable}{@{\extracolsep{\fill}} l c c c c c c @{}}
	\caption{Spin-content weights and nucleon couplings for $X17$.
	Columns list the nuclide, neutron/proton spin weights $\Delta_n,\Delta_p$,
	the spin-fraction-weighted effective couplings for the \emph{hyperfine} sector
	$h_{N}^{\prime(\mathrm{eff})}=h_p^{\prime}\Delta_p+h_n^{\prime}\Delta_n$
	(vector HFS) and $h_{N}^{(\mathrm{eff})}=h_p\Delta_p+h_n\Delta_n$
	(pseudoscalar HFS),
	the coherent coupling $Zh_p^{\prime}+Nh_n^{\prime}$ entering the
	\emph{Lamb shift} via Eq.~\eqref{eq:VX_vector},
	and the experimental nuclear $g$-factor.
	For even-even nuclei with $I=0$, hyperfine-related quantities are not
	applicable and are marked by `—'. The $\mu_{\rm exp}$ values come from PDG
	(light nuclei) and Stone's evaluated tables for stable nuclides.}\\
	\toprule
	Nuclide & $\Delta_n$ & $\Delta_p$
	  & $h_{N}^{\prime(\mathrm{eff})}$
	  & $h_{N}^{(\mathrm{eff})}$
	  & $Zh_p^{\prime}+Nh_n^{\prime}$
	  & $g_N^{(\mathrm{exp})}$ \\
	\midrule
	\endfirsthead
	\midrule
	\endhead
	\midrule
	\endfoot
	\bottomrule
	\endlastfoot

	$^{1}\mathrm{H}$   & $0$              & $1$              & $2.423\times10^{-4}$  & $-2.4\times10^{-3}$   & $2.423\times10^{-4}$  & $5.5856947$ \\
	$^{2}\mathrm{H}$   & $\tfrac{1}{4}$   & $\tfrac{1}{4}$   & $8.176\times10^{-4}$  & $-4.725\times10^{-4}$ & $3.270\times10^{-3}$  & $1.7140255$ \\
	$^{3}\mathrm{H}$   & $0$              & $1$              & $2.423\times10^{-4}$  & $-2.4\times10^{-3}$   & $6.298\times10^{-3}$  & $17.836344$ \\
	$^{3}\mathrm{He}$  & $1$              & $0$              & $3.028\times10^{-3}$  & $5.1\times10^{-4}$    & $3.513\times10^{-3}$  & $-12.736573$ \\
	$^{4}\mathrm{He}$  & —                & —                & —                     & —                     & $6.541\times10^{-3}$  & — \\
	$^{6}\mathrm{Li}$  & $\tfrac{1}{4}$   & $\tfrac{1}{4}$   & $8.176\times10^{-4}$  & $-4.725\times10^{-4}$ & $9.811\times10^{-3}$  & $4.907652$ \\
	$^{7}\mathrm{Li}$  & $0$              & $\tfrac{1}{3}$   & $8.075\times10^{-5}$  & $-8.0\times10^{-4}$   & $1.284\times10^{-2}$  & $15.117825$ \\
	$^{9}\mathrm{Be}$  & $\tfrac{1}{3}$   & $0$              & $1.009\times10^{-3}$  & $1.7\times10^{-4}$    & $1.611\times10^{-2}$  & $-7.021342$ \\
	$^{10}\mathrm{B}$  & $\tfrac{1}{12}$  & $\tfrac{1}{12}$  & $2.725\times10^{-4}$  & $-1.575\times10^{-4}$ & $1.635\times10^{-2}$  & $5.964865$ \\
	$^{11}\mathrm{B}$  & $0$              & $\tfrac{1}{3}$   & $8.075\times10^{-5}$  & $-8.0\times10^{-4}$   & $1.938\times10^{-2}$  & $19.586006$ \\
	$^{12}\mathrm{C}$  & —                & —                & —                     & —                     & $1.962\times10^{-2}$  & — \\
	$^{13}\mathrm{C}$  & $-\tfrac{1}{3}$  & $0$              & $-1.009\times10^{-3}$ & $-1.7\times10^{-4}$   & $2.265\times10^{-2}$  & $18.130871$ \\
	$^{14}\mathrm{N}$  & $-\tfrac{1}{12}$ & $-\tfrac{1}{12}$ & $-2.725\times10^{-4}$ & $1.575\times10^{-4}$  & $2.290\times10^{-2}$  & $5.611513$ \\
	$^{15}\mathrm{N}$  & $0$              & $-\tfrac{1}{3}$  & $-8.075\times10^{-5}$ & $8.0\times10^{-4}$    & $2.592\times10^{-2}$  & $-8.432196$ \\
	$^{16}\mathrm{O}$  & —                & —                & —                     & —                     & $2.616\times10^{-2}$  & — \\
	$^{17}\mathrm{O}$  & $\tfrac{1}{5}$   & $0$              & $6.056\times10^{-4}$  & $1.02\times10^{-4}$   & $2.919\times10^{-2}$  & $-12.780792$ \\
	$^{18}\mathrm{O}$  & —                & —                & —                     & —                     & $3.222\times10^{-2}$  & — \\
	$^{20}\mathrm{Ne}$ & —                & —                & —                     & —                     & $3.270\times10^{-2}$  & — \\
	$^{21}\mathrm{Ne}$ & $-\tfrac{1}{5}$  & $0$              & $-6.056\times10^{-4}$ & $-1.02\times10^{-4}$  & $3.573\times10^{-2}$  & $-9.193117$ \\
	$^{24}\mathrm{Mg}$ & —                & —                & —                     & —                     & $3.924\times10^{-2}$  & — \\
	$^{25}\mathrm{Mg}$ & $\tfrac{1}{5}$   & $0$              & $6.056\times10^{-4}$  & $1.02\times10^{-4}$   & $4.227\times10^{-2}$  & $-8.485658$ \\
	$^{26}\mathrm{Mg}$ & —                & —                & —                     & —                     & $4.530\times10^{-2}$  & — \\
	$^{27}\mathrm{Al}$ & $0$              & $\tfrac{1}{5}$   & $4.846\times10^{-5}$  & $-4.8\times10^{-4}$   & $4.554\times10^{-2}$  & $39.020000$ \\
	$^{28}\mathrm{Si}$ & —                & —                & —                     & —                     & $4.578\times10^{-2}$  & — \\
	$^{29}\mathrm{Si}$ & $1$              & $0$              & $3.028\times10^{-3}$  & $5.1\times10^{-4}$    & $4.881\times10^{-2}$  & $-31.928000$ \\
	$^{30}\mathrm{Si}$ & —                & —                & —                     & —                     & $5.184\times10^{-2}$  & — \\
	$^{31}\mathrm{P}$  & $0$              & $1$              & $2.423\times10^{-4}$  & $-2.4\times10^{-3}$   & $5.208\times10^{-2}$  & $69.605000$ \\
\end{longtable}

%=========================
\section{Vector-$X17$--induced $2S_{1/2}$--$2P_{1/2}$ Lamb shifts in muonic atoms}
\label{app:lamb-vector}
\renewcommand{\thetable}{B\arabic{table}}
\setcounter{table}{0}

\setlength{\tabcolsep}{3pt}
\renewcommand{\arraystretch}{1.06}

\begin{longtable}{@{\extracolsep{\fill}} l cc cc c c c @{}}
	\caption{Vector-$X17$ contribution to the $2S_{1/2}$--$2P_{1/2}$ interval in muonic atoms,
		computed from the coherent potential
		$V_X = \frac{h'_\mu}{4\pi}(Zh_p'+Nh_n')\frac{e^{-m_Xr}}{r}$
		(Eq.~\eqref{eq:VX_vector}).
		Differences shown are
		$\Delta E_{X17}^{(V)} = E_{X17}^{(V)}(2S_{1/2}) - E_{X17}^{(V)}(2P_{1/2})$
		and
		$\Delta E_{L}^{(\mathrm{EM})} = E_{L}^{(\mathrm{EM})}(2S_{1/2}) - E_{L}^{(\mathrm{EM})}(2P_{1/2})$.}
	\label{tab:2S2P_vectorX17_Lamb}\\
	\toprule
	\multirow{2}{*}{Nuclide}
	& \multicolumn{2}{c}{$E_{X17}^{(V)}~[\mathrm{meV}]$}
	& \multicolumn{2}{c}{$E_{L}^{(\mathrm{EM})}~[\mathrm{eV}]$}
	& \multirow{2}{*}{$\Delta E_{X17}^{(V)}~[\mathrm{meV}]$}
	& \multirow{2}{*}{$\Delta E_{L}^{(\mathrm{EM})}~[\mathrm{eV}]$}
	& \multirow{2}{*}{Ratio} \\
	\cmidrule(lr){2-3}\cmidrule(lr){4-5}
	& $2S_{1/2}$ & $2P_{1/2}$
	& $2S_{1/2}$ & $2P_{1/2}$
	& & & {\scriptsize $\Delta E_{X17}^{(V)}/\Delta E_{L}^{(\mathrm{EM})}$} \\
	\midrule
	\endfirsthead
	\midrule
	\endfoot
	\bottomrule
	\endlastfoot
$^{1}\mathrm{H}$   & $2.05125\times 10^{-3}$ & $1.76275\times 10^{-6}$ & $-2.21681\times 10^{-1}$ & $-2.30039\times 10^{-2}$ & $2.04949\times 10^{-3}$ & $-1.98677\times 10^{-1}$ & $-1.03157\times 10^{-5}$ \\
$^{2}\mathrm{H}$   & $3.20481\times 10^{-2}$ & $3.05810\times 10^{-5}$ & $-2.26034\times 10^{-1}$ & $-2.71978\times 10^{-2}$ & $3.20175\times 10^{-2}$ & $-1.98837\times 10^{-1}$ & $-1.61024\times 10^{-4}$ \\
$^{3}\mathrm{H}$   & $6.49607\times 10^{-2}$ & $6.42143\times 10^{-5}$ & $-2.43984\times 10^{-1}$ & $-2.88209\times 10^{-2}$ & $6.48965\times 10^{-2}$ & $-2.15163\times 10^{-1}$ & $-3.01615\times 10^{-4}$ \\
$^{3}\mathrm{He}$  & $2.45829\times 10^{-1}$ & $9.72451\times 10^{-4}$ & $-3.64950\times 10^{0}$  & $-5.69311\times 10^{-1}$ & $2.44856\times 10^{-1}$ & $-3.08019\times 10^{0}$  & $-7.94940\times 10^{-5}$ \\
$^{4}\mathrm{He}$  & $4.69856\times 10^{-1}$ & $1.88829\times 10^{-3}$ & $-1.97317\times 10^{0}$  & $-5.84337\times 10^{-1}$ & $4.67968\times 10^{-1}$ & $-1.38883\times 10^{0}$  & $-3.36951\times 10^{-4}$ \\
$^{6}\mathrm{Li}$  & $2.07425\times 10^{0}$  & $1.91350\times 10^{-2}$ & $-4.54573\times 10^{0}$  & $-2.99490\times 10^{0}$  & $2.05511\times 10^{0}$  & $-1.55083\times 10^{0}$  & $-1.32517\times 10^{-3}$ \\
$^{7}\mathrm{Li}$  & $2.73512\times 10^{0}$  & $2.53443\times 10^{-2}$ & $-4.72410\times 10^{0}$  & $-3.01603\times 10^{0}$  & $2.70977\times 10^{0}$  & $-1.70807\times 10^{0}$  & $-1.58645\times 10^{-3}$ \\
$^{9}\mathrm{Be}$  & $7.04029\times 10^{0}$  & $1.16646\times 10^{-1}$ & $-2.21206\times 10^{1}$  & $-9.01208\times 10^{0}$  & $6.92365\times 10^{0}$  & $-1.31086\times 10^{1}$  & $-5.28177\times 10^{-4}$ \\
$^{10}\mathrm{B}$  & $1.21548\times 10^{1}$  & $3.13017\times 10^{-1}$ & $-2.15338\times 10^{1}$  & $-2.04440\times 10^{1}$  & $1.18417\times 10^{1}$  & $-1.08981\times 10^{0}$  & $-1.08658\times 10^{-2}$ \\
$^{11}\mathrm{B}$  & $1.44337\times 10^{1}$  & $3.72639\times 10^{-1}$ & $-1.48651\times 10^{1}$  & $-2.04965\times 10^{1}$  & $1.40610\times 10^{1}$  & $5.63135\times 10^{0}$   & $2.49692\times 10^{-3}$ \\
$^{12}\mathrm{C}$  & $2.20654\times 10^{1}$  & $8.16132\times 10^{-1}$ & $-3.22973\times 10^{1}$  & $-3.95110\times 10^{1}$  & $2.12493\times 10^{1}$  & $7.21377\times 10^{0}$   & $2.94566\times 10^{-3}$ \\
$^{13}\mathrm{C}$  & $2.54461\times 10^{1}$ & $9.44980\times 10^{-1}$ & $-1.56057\times 10^{2}$ & $-3.96263\times 10^{1}$ & $2.45011\times 10^{1}$ & $-1.16430\times 10^{2}$ & $-2.10436\times 10^{-4}$ \\
$^{14}\mathrm{N}$  & $3.60886\times 10^{1}$  & $1.80258\times 10^{0}$  & $-4.85038\times 10^{2}$  & $-6.85058\times 10^{1}$  & $3.42860\times 10^{1}$  & $-4.16532\times 10^{2}$  & $-8.23130\times 10^{-5}$ \\
$^{15}\mathrm{N}$  & $4.08473\times 10^{1}$  & $6.90205\times 10^{-1}$ & $-4.96745\times 10^{2}$  & $-2.04109\times 10^{1}$  & $4.01571\times 10^{1}$  & $-4.76334\times 10^{2}$  & $-8.43044\times 10^{-5}$ \\
$^{16}\mathrm{O}$  & $5.39722\times 10^{1}$  & $3.52607\times 10^{0}$  & $-1.97357\times 10^{2}$  & $-1.11191\times 10^{2}$  & $5.04461\times 10^{1}$  & $-8.61668\times 10^{1}$  & $-5.85448\times 10^{-4}$ \\
$^{17}\mathrm{O}$  & $6.02308\times 10^{1}$  & $3.94096\times 10^{0}$  & $-2.19007\times 10^{2}$  & $-1.11192\times 10^{2}$  & $5.62898\times 10^{1}$  & $-1.07815\times 10^{2}$  & $-5.22094\times 10^{-4}$ \\
$^{18}\mathrm{O}$  & $6.64632\times 10^{1}$  & $4.35638\times 10^{0}$  & $-2.48024\times 10^{2}$  & $-1.11202\times 10^{2}$  & $6.21069\times 10^{1}$  & $-1.36822\times 10^{2}$  & $-4.53925\times 10^{-4}$ \\
$^{20}\mathrm{Ne}$ & $1.04646\times 10^{2}$  & $1.04895\times 10^{1}$  & $-4.98728\times 10^{3}$  & $-2.42051\times 10^{2}$  & $9.41570\times 10^{1}$  & $-4.74523\times 10^{3}$  & $-1.98424\times 10^{-5}$ \\
$^{21}\mathrm{Ne}$ & $1.14390\times 10^{2}$  & $1.14731\times 10^{1}$  & $-4.63871\times 10^{3}$  & $-2.42193\times 10^{2}$  & $1.02917\times 10^{2}$  & $-4.39652\times 10^{3}$  & $-2.34088\times 10^{-5}$ \\
$^{24}\mathrm{Mg}$ & $1.74876\times 10^{2}$  & $2.47678\times 10^{1}$  & $-7.29154\times 10^{3}$  & $-4.62356\times 10^{2}$  & $1.50109\times 10^{2}$  & $-6.82919\times 10^{3}$  & $-2.19805\times 10^{-5}$ \\
$^{25}\mathrm{Mg}$ & $1.88388\times 10^{2}$  & $2.66985\times 10^{1}$  & $-7.12773\times 10^{3}$  & $-4.62545\times 10^{2}$  & $1.61689\times 10^{2}$  & $-6.66518\times 10^{3}$  & $-2.42588\times 10^{-5}$ \\
$^{26}\mathrm{Mg}$ & $2.02208\times 10^{2}$  & $2.86302\times 10^{1}$  & $-6.50594\times 10^{3}$  & $-4.62719\times 10^{2}$  & $1.73578\times 10^{2}$  & $-6.04322\times 10^{3}$  & $-2.87227\times 10^{-5}$ \\
$^{27}\mathrm{Al}$ & $2.33870\times 10^{2}$  & $3.83667\times 10^{1}$  & $-1.73382\times 10^{4}$  & $-6.32364\times 10^{2}$  & $1.95503\times 10^{2}$  & $-1.67058\times 10^{4}$  & $-1.17027\times 10^{-5}$ \\
$^{28}\mathrm{Si}$ & $2.03453\times 10^{2}$  & $2.89659\times 10^{1}$  & $-7.67313\times 10^{3}$  & $-4.63034\times 10^{2}$  & $1.74487\times 10^{2}$  & $-7.21009\times 10^{3}$  & $-2.42004\times 10^{-5}$ \\
$^{29}\mathrm{Si}$ & $2.90383\times 10^{2}$  & $5.33832\times 10^{1}$  & $-8.84039\times 10^{4}$  & $-8.07264\times 10^{2}$  & $2.37000\times 10^{2}$  & $-8.75966\times 10^{4}$  & $-2.70558\times 10^{-6}$ \\
$^{30}\mathrm{Si}$ & $2.30301\times 10^{2}$  & $3.28321\times 10^{1}$  & $-7.57372\times 10^{3}$  & $-4.63304\times 10^{2}$  & $1.97469\times 10^{2}$  & $-7.11042\times 10^{3}$  & $-2.77718\times 10^{-5}$ \\
$^{31}\mathrm{P}$  & $3.36263\times 10^{2}$ & $7.22637\times 10^{1}$ & $-3.74113\times 10^{3}$ & $-1.20281\times 10^{3}$ & $2.63999\times 10^{2}$ & $-2.53832\times 10^{3}$ & $-1.04005\times 10^{-4}$ \\
\end{longtable}

%=========================
\section{Vector-$X17$-induced hyperfine structure in muonic atoms}
\label{app:hfs-vector}
\renewcommand{\thetable}{C\arabic{table}}
\setcounter{table}{0}

\subsection{$1S_{1/2}$ hyperfine structure}
\label{app:hfs-vector:1s}

\setlength{\tabcolsep}{3pt}
\renewcommand{\arraystretch}{1.06}

\begin{longtable}{@{\extracolsep{\fill}} l cc cc c c c @{}}
\caption{Vector $X17$ in $1S_{1/2}$: level energies for non--even-even nuclei.
Subcolumns are $F=S_N\!-\!J$ and $F=S_N\!+\!J$ (for $1S_{1/2}$, $J=\tfrac12$).
Differences shown are $\Delta E_{X17}$ and $\Delta E_{\mathrm{EM}}$.}
\label{tab:1S_vectorX17_all}\\

\toprule
\multirow{2}{*}{Nuclide}
& \multicolumn{2}{c}{$E_{X17}^{(V)}~[\mathrm{meV}]$}
& \multicolumn{2}{c}{$E_{\mathrm{HFS}}^{(\mathrm{EM})}~[\mathrm{eV}]$}
& \multirow{2}{*}{$\Delta E_{X17}^{(V)}~[\mathrm{meV}]$}
& \multirow{2}{*}{$\Delta E_{\mathrm{HFS}}^{(\mathrm{EM})}~[\mathrm{eV}]$}
& \multirow{2}{*}{Ratio} \\
\cmidrule(lr){2-3}\cmidrule(lr){4-5}
& $F=S_N\!-\!J$ & $F=S_N\!+\!J$
& $F=S_N\!-\!J$ & $F=S_N\!+\!J$
& & & {\scriptsize $\Delta E_{X17}^{(V)}/\Delta E_{\mathrm{HFS}}^{(\mathrm{EM})}$} \\
\midrule
\endfirsthead
\midrule
\endfoot
\bottomrule
\endlastfoot

$^{1}\mathrm{H}$  & $1.64795\times 10^{-2}$ & $1.64538\times 10^{-2}$ & $-1.37650\times 10^{-1}$ & $4.55845\times 10^{-2}$ & $-2.56779\times 10^{-5}$ & $1.83235\times 10^{-1}$ & $-1.40136\times 10^{-7}$ \\
$^{2}\mathrm{H}$  & $6.43445\times 10^{-2}$ & $6.43132\times 10^{-2}$ & $-3.29861\times 10^{-2}$ & $1.66585\times 10^{-2}$ & $-3.13625\times 10^{-5}$ & $4.96446\times 10^{-2}$ & $-6.31741\times 10^{-7}$ \\
$^{3}\mathrm{H}$  & $2.00836\times 10^{-2}$ & $2.00545\times 10^{-2}$ & $-1.79037\times 10^{-1}$ & $5.98639\times 10^{-2}$ & $-2.91580\times 10^{-5}$ & $2.38901\times 10^{-1}$ & $-1.22051\times 10^{-7}$ \\
$^{3}\mathrm{He}$ & $1.71420\times 10^{0}$  & $1.71584\times 10^{0}$  & $1.03392\times 10^{0}$  & $-3.43674\times 10^{-1}$ & $1.63759\times 10^{-3}$ & $-1.37759\times 10^{0}$ & $-1.18873\times 10^{-6}$ \\
$^{6}\mathrm{Li}$ & $1.41133\times 10^{0}$  & $1.41038\times 10^{0}$  & $-8.91509\times 10^{-1}$ & $4.44334\times 10^{-1}$ & $-9.51181\times 10^{-4}$ & $1.33584\times 10^{0}$ & $-7.12046\times 10^{-7}$ \\
$^{7}\mathrm{Li}$ & $1.40571\times 10^{0}$  & $1.40308\times 10^{0}$  & $-2.99664\times 10^{0}$  & $1.76060\times 10^{0}$ & $-2.62341\times 10^{-3}$ & $4.75724\times 10^{0}$ & $-5.51411\times 10^{-7}$ \\
$^{9}\mathrm{Be}$ & $3.64731\times 10^{0}$  & $3.64770\times 10^{0}$  & $2.88950\times 10^{0}$  & $-1.71041\times 10^{0}$ & $3.84843\times 10^{-4}$ & $-4.59991\times 10^{0}$ & $-8.36672\times 10^{-8}$ \\
$^{10}\mathrm{B}$ & $1.70084\times 10^{0}$  & $1.69689\times 10^{0}$  & $-9.20307\times 10^{0}$  & $6.92823\times 10^{0}$ & $-3.94983\times 10^{-3}$ & $1.61313\times 10^{1}$ & $-2.44823\times 10^{-7}$ \\
$^{11}\mathrm{B}$ & $5.05247\times 10^{-1}$ & $5.04533\times 10^{-1}$ & $-1.67214\times 10^{1}$ & $1.08051\times 10^{1}$ & $-7.13596\times 10^{-4}$ & $2.75264\times 10^{1}$ & $-2.59215\times 10^{-8}$ \\
$^{13}\mathrm{C}$ & $-9.61119\times 10^{0}$ & $-9.60465\times 10^{0}$ & $-9.51785\times 10^{0}$ & $3.22829\times 10^{0}$ & $6.53708\times 10^{-3}$ & $1.27461\times 10^{1}$ & $5.12922\times 10^{-7}$ \\
$^{14}\mathrm{N}$ & $-3.65623\times 10^{0}$ & $-3.65430\times 10^{0}$ & $-5.37053\times 10^{0}$ & $2.70519\times 10^{0}$ & $1.93417\times 10^{-3}$ & $8.07572\times 10^{0}$ & $2.39494\times 10^{-7}$ \\
$^{15}\mathrm{N}$ & $-1.08190\times 10^{0}$ & $-1.08308\times 10^{0}$ & $5.03209\times 10^{0}$  & $-1.90055\times 10^{0}$ & $-1.17717\times 10^{-3}$ & $-6.93264\times 10^{0}$ & $1.69774\times 10^{-7}$ \\
$^{17}\mathrm{O}$ & $1.10691\times 10^{1}$  & $1.10975\times 10^{1}$  & $1.05051\times 10^{2}$  & $-7.49520\times 10^{1}$ & $2.83746\times 10^{-2}$ & $-1.79903\times 10^{2}$ & $-1.57696\times 10^{-7}$ \\
$^{21}\mathrm{Ne}$ & $-1.69286\times 10^{1}$ & $-1.69378\times 10^{1}$ & $1.86558\times 10^{1}$  & $-1.02988\times 10^{1}$ & $-9.16707\times 10^{-3}$ & $-2.89545\times 10^{1}$ & $3.16603\times 10^{-7}$ \\
$^{25}\mathrm{Mg}$ & $2.38868\times 10^{1}$  & $2.38581\times 10^{1}$  & $4.25753\times 10^{1}$  & $-3.55384\times 10^{1}$ & $-2.87315\times 10^{-2}$ & $-7.81136\times 10^{1}$ & $3.67827\times 10^{-7}$ \\
$^{27}\mathrm{Al}$ & $2.22602\times 10^{0}$  & $2.25491\times 10^{0}$  & $-1.47751\times 10^{2}$ & $7.62922\times 10^{2}$ & $2.88963\times 10^{-2}$ & $9.10674\times 10^{2}$ & $3.17235\times 10^{-8}$ \\
$^{29}\mathrm{Si}$ & $1.57020\times 10^{2}$  & $1.56377\times 10^{2}$  & $7.44958\times 10^{1}$  & $-3.22391\times 10^{1}$ & $-6.43146\times 10^{-1}$ & $-1.06735\times 10^{2}$ & $6.02570\times 10^{-6}$ \\
$^{31}\mathrm{P}$  & $1.37586\times 10^{1}$  & $1.38510\times 10^{1}$  & $-6.07764\times 10^{2}$ & $1.29332\times 10^{2}$ & $9.23910\times 10^{-2}$ & $7.37096\times 10^{2}$ & $1.25331\times 10^{-7}$ \\
\end{longtable}

\subsection{$2P_{1/2}$ hyperfine structure}
\label{app:hfs-vector:2p}

\begingroup
\footnotesize
\setlength{\tabcolsep}{2.0pt}
\renewcommand{\arraystretch}{1.03}
\begin{longtable}{@{\extracolsep{\fill}} l cc cc c c c @{}}
\caption{Vector $X17$ in $2P_{1/2}$: level energies for non--even-even nuclei.}
\label{tab:2P_vectorX17_all}\\

\toprule
\multirow{2}{*}{Nuclide}
& \multicolumn{2}{c}{$E_{X17}^{(V)}~[\mathrm{meV}]$}
& \multicolumn{2}{c}{$E_{\mathrm{HFS}}^{(\mathrm{EM})}~[\mathrm{eV}]$}
& \multirow{2}{*}{$\Delta E_{X17}^{(V)}~[\mathrm{meV}]$}
& \multirow{2}{*}{$\Delta E_{\mathrm{HFS}}^{(\mathrm{EM})}~[\mathrm{eV}]$}
& \multirow{2}{*}{Ratio} \\
\cmidrule(lr){2-3}\cmidrule(lr){4-5}
& $F=S_N\!-\!J$ & $F=S_N\!+\!J$
& $F=S_N\!-\!J$ & $F=S_N\!+\!J$
& & & {\scriptsize $\Delta E_{X17}^{(V)}/\Delta E_{\mathrm{HFS}}^{(\mathrm{EM})}$} \\
\midrule
\endfirsthead
\midrule
\endfoot
\bottomrule
\endlastfoot

$^{1}\mathrm{H}$ & $1.76523\times 10^{-6}$ & $1.76193\times 10^{-6}$ & $-1.40533\times 10^{-3}$ & $4.68424\times 10^{-4}$ & $-3.30573\times 10^{-9}$ & $1.87376\times 10^{-3}$ & $-1.76422\times 10^{-9}$ \\
$^{2}\mathrm{H}$ & $7.65256\times 10^{-6}$ & $7.64300\times 10^{-6}$ & $-3.47656\times 10^{-4}$ & $1.73826\times 10^{-4}$ & $-9.55542\times 10^{-9}$ & $5.21482\times 10^{-4}$ & $-1.83229\times 10^{-8}$ \\
$^{3}\mathrm{H}$ & $2.47196\times 10^{-6}$ & $2.47000\times 10^{-6}$ & $-1.93300\times 10^{-3}$ & $6.44301\times 10^{-4}$ & $-1.96472\times 10^{-9}$ & $2.57730\times 10^{-3}$ & $-7.62442\times 10^{-10}$ \\
$^{3}\mathrm{He}$ & $8.38409\times 10^{-4}$ & $8.38129\times 10^{-4}$ & $1.39667\times 10^{-2}$ & $-4.65586\times 10^{-3}$ & $-2.79382\times 10^{-7}$ & $-1.86226\times 10^{-2}$ & $1.50034\times 10^{-8}$ \\
$^{6}\mathrm{Li}$ & $1.59510\times 10^{-3}$ & $1.59430\times 10^{-3}$ & $-1.31292\times 10^{-2}$ & $6.56444\times 10^{-3}$ & $-7.99141\times 10^{-7}$ & $1.96936\times 10^{-2}$ & $-4.05871\times 10^{-8}$ \\
$^{7}\mathrm{Li}$ & $1.59460\times 10^{-3}$ & $1.59342\times 10^{-3}$ & $-4.36932\times 10^{-2}$ & $2.62136\times 10^{-2}$ & $-1.18290\times 10^{-6}$ & $6.99068\times 10^{-2}$ & $-1.69200\times 10^{-8}$ \\
$^{9}\mathrm{Be}$ & $7.30729\times 10^{-3}$ & $7.30507\times 10^{-3}$ & $3.79765\times 10^{-2}$ & $-2.27869\times 10^{-2}$ & $-2.21743\times 10^{-6}$ & $-6.07634\times 10^{-2}$ & $3.64978\times 10^{-8}$ \\
$^{10}\mathrm{B}$ & $5.21933\times 10^{-3}$ & $5.21513\times 10^{-3}$ & $-9.12873\times 10^{-2}$ & $6.84604\times 10^{-2}$ & $-4.19645\times 10^{-6}$ & $1.59748\times 10^{-1}$ & $-2.62663\times 10^{-8}$ \\
$^{11}\mathrm{B}$ & $1.55321\times 10^{-3}$ & $1.55230\times 10^{-3}$ & $-1.70912\times 10^{-1}$ & $1.02534\times 10^{-1}$ & $-9.09914\times 10^{-7}$ & $2.73446\times 10^{-1}$ & $-3.32723\times 10^{-9}$ \\
$^{13}\mathrm{C}$ & $-4.21071\times 10^{-2}$ & $-4.20964\times 10^{-2}$ & $-1.39917\times 10^{-1}$ & $4.66361\times 10^{-2}$ & $1.06944\times 10^{-5}$ & $1.86553\times 10^{-1}$ & $5.73169\times 10^{-8}$ \\
$^{14}\mathrm{N}$ & $-2.14525\times 10^{-2}$ & $-2.14468\times 10^{-2}$ & $-8.54219\times 10^{-2}$ & $4.27106\times 10^{-2}$ & $5.69459\times 10^{-6}$ & $1.28133\times 10^{-1}$ & $4.44482\times 10^{-8}$ \\
$^{15}\mathrm{N}$ & $-6.37011\times 10^{-3}$ & $-6.36952\times 10^{-3}$ & $9.00132\times 10^{-2}$ & $-3.00064\times 10^{-2}$ & $5.85995\times 10^{-10}$ & $-1.20020\times 10^{-1}$ & $-4.88209\times 10^{-12}$ \\
$^{17}\mathrm{O}$ & $8.17780\times 10^{-2}$ & $8.17614\times 10^{-2}$ & $4.21766\times 10^{-1}$ & $-3.01299\times 10^{-1}$ & $-1.65817\times 10^{-5}$ & $-7.23065\times 10^{-1}$ & $2.29327\times 10^{-8}$ \\
$^{21}\mathrm{Ne}$ & $-1.94430\times 10^{-1}$ & $-1.94391\times 10^{-1}$ & $3.45176\times 10^{-1}$ & $-2.06942\times 10^{-1}$ & $3.89972\times 10^{-5}$ & $-5.52118\times 10^{-1}$ & $-7.06301\times 10^{-8}$ \\
$^{25}\mathrm{Mg}$ & $3.82583\times 10^{-1}$ & $3.82507\times 10^{-1}$ & $6.57331\times 10^{-1}$ & $-4.69650\times 10^{-1}$ & $-7.64737\times 10^{-5}$ & $-1.12698\times 10^{0}$ & $6.78643\times 10^{-8}$ \\
$^{27}\mathrm{Al}$ & $4.08235\times 10^{-2}$ & $4.07973\times 10^{-2}$ & $-3.56816\times 10^{0}$ & $2.55029\times 10^{0}$ & $-2.61560\times 10^{-5}$ & $6.11845\times 10^{0}$ & $-4.27545\times 10^{-9}$ \\
$^{29}\mathrm{Si}$ & $3.31316\times 10^{0}$ & $3.31403\times 10^{0}$ & $1.47523\times 10^{0}$ & $-4.93339\times 10^{-1}$ & $8.70304\times 10^{-4}$ & $-1.96857\times 10^{0}$ & $-4.42102\times 10^{-7}$ \\
$^{31}\mathrm{P}$ & $3.36530\times 10^{-1}$ & $3.36420\times 10^{-1}$ & $-3.69986\times 10^{0}$ & $1.23304\times 10^{0}$ & $-1.09336\times 10^{-4}$ & $4.93291\times 10^{0}$ & $-2.21644\times 10^{-8}$ \\
\end{longtable}
\endgroup

%=========================
\section{Pseudoscalar-$X17$-induced hyperfine structure in muonic atoms}
\label{app:hfs-ps}
\renewcommand{\thetable}{D\arabic{table}}
\setcounter{table}{0}

\subsection{$1S_{1/2}$ hyperfine structure}
\label{app:hfs-ps:1s}

\begin{longtable}{@{\extracolsep{\fill}} l cc cc c c c @{}}
\caption{Pseudoscalar $X17$: level energies for non--even-even nuclei.
Subcolumns are $F=S_N\!-\!J$ and $F=S_N\!+\!J$ (for $1S_{1/2}$, $J=\tfrac12$).}\label{tab:1S_PseudoscalarX17_alll}\\

\toprule
\multirow{2}{*}{Nuclide}
& \multicolumn{2}{c}{$E_{X17}^{(A)}\,[\mathrm{meV}]$}
& \multicolumn{2}{c}{$E_{\mathrm{HFS}}^{(\mathrm{EM})}\,[\mathrm{eV}]$}
& \multirow{2}{*}{$\Delta E_{X17}\,[\mathrm{meV}]$}
& \multirow{2}{*}{$\Delta E_{\mathrm{EM}}\,[\mathrm{eV}]$}
& \multirow{2}{*}{Ratio} \\
\cmidrule(lr){2-3}\cmidrule(lr){4-5}
& $F=S_N\!-\!J$ & $F=S_N\!+\!J$ & $F=S_N\!-\!J$ & $F=S_N\!+\!J$ & & & \\
\midrule
\endfirsthead
\midrule
\endfoot
\bottomrule
\endlastfoot

$^{1}\mathrm{H}$  & $-1.62898\times 10^{-4}$ & $5.23107\times 10^{-5}$ & $-1.37796\times 10^{-1}$ & $4.56325\times 10^{-2}$ & $2.15208\times 10^{-4}$ & $1.83428\times 10^{-1}$ & $1.17326\times 10^{-6}$ \\
$^{2}\mathrm{H}$  & $-1.92273\times 10^{-5}$ & $1.04034\times 10^{-5}$ & $-3.15703\times 10^{-2}$ & $1.59376\times 10^{-2}$ & $2.96307\times 10^{-5}$ & $4.75079\times 10^{-2}$ & $6.23710\times 10^{-7}$ \\
$^{3}\mathrm{H}$  & $-6.48738\times 10^{-5}$ & $2.22504\times 10^{-5}$ & $-1.78672\times 10^{-1}$ & $5.97417\times 10^{-2}$ & $8.71242\times 10^{-5}$ & $2.38414\times 10^{-1}$ & $3.65452\times 10^{-7}$ \\
$^{3}\mathrm{He}$ & $2.08131\times 10^{-4}$  & $-6.84764\times 10^{-5}$ & $1.02733\times 10^{0}$  & $-3.41520\times 10^{-1}$ & $-2.76607\times 10^{-4}$ & $-1.36885\times 10^{0}$ & $2.02053\times 10^{-7}$ \\
$^{6}\mathrm{Li}$ & $-5.46531\times 10^{-4}$ & $2.70768\times 10^{-4}$ & $-8.91760\times 10^{-1}$ & $4.44348\times 10^{-1}$ & $8.17298\times 10^{-4}$ & $1.33611\times 10^{0}$ & $6.11567\times 10^{-7}$ \\
$^{7}\mathrm{Li}$ & $-8.03830\times 10^{-4}$ & $4.02161\times 10^{-4}$ & $-2.75675\times 10^{0}$ & $1.56991\times 10^{0}$ & $1.20699\times 10^{-3}$ & $4.32666\times 10^{0}$ & $2.78923\times 10^{-7}$ \\
$^{9}\mathrm{Be}$ & $5.56011\times 10^{-4}$  & $-3.33877\times 10^{-4}$ & $2.69550\times 10^{0}$ & $-1.61844\times 10^{0}$ & $-8.89888\times 10^{-4}$ & $-4.31393\times 10^{0}$ & $2.06329\times 10^{-7}$ \\
$^{10}\mathrm{B}$ & $1.28740\times 10^{-2}$  & $-3.49457\times 10^{-3}$ & $-1.84258\times 10^{1}$ & $6.45649\times 10^{0}$ & $-1.63685\times 10^{-2}$ & $2.48823\times 10^{1}$ & $6.57740\times 10^{-7}$ \\
$^{11}\mathrm{B}$ & $-8.87346\times 10^{-3}$ & $5.07868\times 10^{-3}$ & $-1.75574\times 10^{1}$ & $1.02008\times 10^{1}$ & $1.39521\times 10^{-2}$ & $2.77582\times 10^{1}$ & $5.02488\times 10^{-7}$ \\
$^{13}\mathrm{C}$ & $-9.66528\times 10^{-4}$ & $3.25034\times 10^{-4}$ & $-9.72284\times 10^{0}$ & $3.25666\times 10^{0}$ & $1.29156\times 10^{-3}$ & $1.29795\times 10^{1}$ & $9.95052\times 10^{-8}$ \\
$^{14}\mathrm{N}$ & $7.86014\times 10^{-4}$  & $-7.46216\times 10^{-4}$ & $-3.74528\times 10^{0}$ & $2.55625\times 10^{0}$ & $-1.53223\times 10^{-3}$ & $6.30153\times 10^{0}$ & $2.43208\times 10^{-7}$ \\
$^{15}\mathrm{N}$ & $-5.27350\times 10^{-3}$ & $1.40266\times 10^{-3}$ & $5.35887\times 10^{0}$ & $-1.58260\times 10^{0}$ & $6.67616\times 10^{-3}$ & $-6.94147\times 10^{0}$ & $9.61906\times 10^{-7}$ \\
$^{17}\mathrm{O}$ & $6.76740\times 10^{-4}$  & $-3.56875\times 10^{-3}$ & $1.43279\times 10^{1}$ & $-3.12414\times 10^{1}$ & $-4.24549\times 10^{-3}$ & $-4.55692\times 10^{1}$ & $9.31548\times 10^{-8}$ \\
$^{21}\mathrm{Ne}$& $-5.18206\times 10^{-3}$ & $2.89136\times 10^{-3}$ & $1.90020\times 10^{1}$ & $-1.03316\times 10^{1}$ & $8.07342\times 10^{-3}$ & $-2.93336\times 10^{1}$ & $2.75187\times 10^{-7}$ \\
$^{25}\mathrm{Mg}$& $4.43419\times 10^{-3}$  & $-7.05199\times 10^{-3}$ & $2.97668\times 10^{1}$ & $-3.87654\times 10^{1}$ & $-1.14862\times 10^{-2}$ & $-6.85322\times 10^{1}$ & $1.67594\times 10^{-7}$ \\
$^{27}\mathrm{Al}$& $-2.85227\times 10^{-2}$ & $2.45572\times 10^{-2}$ & $-1.73896\times 10^{2}$ & $1.42511\times 10^{2}$ & $5.30799\times 10^{-2}$ & $3.16407\times 10^{2}$ & $1.67780\times 10^{-7}$ \\
$^{29}\mathrm{Si}$& $2.36601\times 10^{-2}$  & $-1.02860\times 10^{-2}$ & $9.80510\times 10^{1}$ & $-4.08207\times 10^{1}$ & $-3.39461\times 10^{-2}$ & $-1.38872\times 10^{2}$ & $2.44485\times 10^{-7}$ \\
$^{31}\mathrm{P}$ & $-7.07436\times 10^{-1}$ & $2.17207\times 10^{-3}$ & $-1.15398\times 10^{3}$ & $2.16058\times 10^{0}$ & $7.09608\times 10^{-1}$ & $1.15614\times 10^{3}$ & $6.13708\times 10^{-7}$ \\
\end{longtable}

\subsection{$2P_{1/2}$ hyperfine structure}
\label{app:hfs-ps:2p}

% --- compact spacing (only affects this table) ---
\setlength{\tabcolsep}{3pt}
\renewcommand{\arraystretch}{1.06}

\begin{longtable}{@{\extracolsep{\fill}} l cc cc c c c @{}}
\caption{Pseudoscalar $X17$ in $2P_{1/2}$: level energies for non--even-even nuclei.
Subcolumns are $F=S_N\!-\!J$ and $F=S_N\!+\!J$ (for $2P_{1/2}$, $J=\tfrac12$).
Differences shown are $\Delta E_{X17}$ and $\Delta E_{\mathrm{EM}}$.}
\label{tab:2P_PseudoscalarX17_all}\\

\toprule
\multirow{2}{*}{Nuclide}
& \multicolumn{2}{c}{$E_{X17}^{(A)}~[\mu\mathrm{eV}]$}
& \multicolumn{2}{c}{$E_{\mathrm{HFS}}^{(\mathrm{EM})}~[\mathrm{eV}]$}
& \multirow{2}{*}{$\Delta E_{X17}~[\mu\mathrm{eV}]$}
& \multirow{2}{*}{$\Delta E_{\mathrm{HFS}}^{(\mathrm{EM})}~[\mathrm{eV}]$}
& \multirow{2}{*}{Ratio} \\
\cmidrule(lr){2-3}\cmidrule(lr){4-5}
& $F=S_N\!-\!J$ & $F=S_N\!+\!J$
& $F=S_N\!-\!J$ & $F=S_N\!+\!J$
& & & {\scriptsize $\Delta E_{X17}^{(A)}/\Delta E_{\mathrm{HFS}}^{(\mathrm{EM})}$} \\
\midrule
\endfirsthead

\midrule
\endfoot

\bottomrule
\endlastfoot

$^{1}\mathrm{H}$  & $-7.20457\times 10^{-6}$ & $2.40640\times 10^{-6}$ & $-1.40533\times 10^{-3}$ & $4.68424\times 10^{-4}$ & $9.61097\times 10^{-6}$ & $1.87376\times 10^{-3}$ & $5.12925\times 10^{-9}$ \\
$^{2}\mathrm{H}$  & $-4.67867\times 10^{-3}$ & $-4.67258\times 10^{-3}$ & $-3.51268\times 10^{-4}$ & $1.75632\times 10^{-4}$ & $6.09541\times 10^{-6}$ & $5.26900\times 10^{-4}$ & $1.15679\times 10^{-8}$ \\
$^{3}\mathrm{H}$  & $-3.24857\times 10^{-6}$ & $1.08512\times 10^{-6}$ & $-1.93300\times 10^{-3}$ & $6.44301\times 10^{-4}$ & $4.33369\times 10^{-6}$ & $2.57730\times 10^{-3}$ & $1.68159\times 10^{-9}$ \\
$^{3}\mathrm{He}$ & $7.56931\times 10^{-6}$ & $-2.51092\times 10^{-6}$ & $1.39667\times 10^{-2}$ & $-4.65586\times 10^{-3}$ & $-1.00802\times 10^{-5}$ & $-1.86226\times 10^{-2}$ & $5.41320\times 10^{-10}$ \\
$^{6}\mathrm{Li}$ & $1.61007\times 10^{-5}$ & $-8.05892\times 10^{-6}$ & $-1.31292\times 10^{-2}$ & $6.56443\times 10^{-3}$ & $-2.41596\times 10^{-5}$ & $1.96936\times 10^{-2}$ & $1.22676\times 10^{-9}$ \\
$^{7}\mathrm{Li}$ & $3.35391\times 10^{-5}$ & $-2.00607\times 10^{-5}$ & $-4.36932\times 10^{-2}$ & $2.62136\times 10^{-2}$ & $-5.35998\times 10^{-5}$ & $6.99068\times 10^{-2}$ & $7.66794\times 10^{-10}$ \\
$^{9}\mathrm{Be}$ & $-8.22068\times 10^{-5}$ & $4.93810\times 10^{-5}$ & $3.79765\times 10^{-2}$ & $-2.27869\times 10^{-2}$ & $1.31588\times 10^{-4}$ & $-6.07634\times 10^{-2}$ & $-2.16578\times 10^{-9}$ \\
$^{10}\mathrm{B}$ & $-8.69145\times 10^{-4}$ & $6.51198\times 10^{-4}$ & $-9.12873\times 10^{-2}$ & $6.84604\times 10^{-2}$ & $1.52034\times 10^{-3}$ & $1.59748\times 10^{-1}$ & $9.51807\times 10^{-9}$ \\
$^{11}\mathrm{B}$ & $1.46153\times 10^{-3}$ & $-8.75433\times 10^{-4}$ & $-1.70912\times 10^{-1}$ & $1.02534\times 10^{-1}$ & $-2.33697\times 10^{-3}$ & $2.73446\times 10^{-1}$ & $8.54500\times 10^{-9}$ \\
$^{13}\mathrm{C}$ & $5.03453\times 10^{-4}$ & $-1.67676\times 10^{-4}$ & $-1.39918\times 10^{-1}$ & $4.66361\times 10^{-2}$ & $-6.71129\times 10^{-4}$ & $1.86554\times 10^{-1}$ & $-3.59774\times 10^{-9}$ \\
$^{14}\mathrm{N}$ & $-1.36603\times 10^{-3}$ & $6.81908\times 10^{-4}$ & $-8.54247\times 10^{-2}$ & $4.27093\times 10^{-2}$ & $2.04794\times 10^{-3}$ & $1.28134\times 10^{-1}$ & $1.59790\times 10^{-8}$ \\
$^{15}\mathrm{N}$ & $1.81513\times 10^{-3}$ & $-6.04287\times 10^{-4}$ & $2.66533\times 10^{-2}$ & $-8.88422\times 10^{-3}$ & $-2.41942\times 10^{-3}$ & $-3.55375\times 10^{-2}$ & $6.80751\times 10^{-8}$ \\
$^{17}\mathrm{O}$ & $-2.76768\times 10^{-3}$ & $1.97815\times 10^{-3}$ & $4.21767\times 10^{-1}$ & $-3.01300\times 10^{-1}$ & $4.74582\times 10^{-3}$ & $-7.23067\times 10^{-1}$ & $-6.56434\times 10^{-9}$ \\
$^{21}\mathrm{Ne}$& $5.12853\times 10^{-3}$ & $-3.11381\times 10^{-3}$ & $3.45073\times 10^{-1}$ & $-2.07226\times 10^{-1}$ & $-8.24233\times 10^{-3}$ & $-5.52298\times 10^{-1}$ & $1.49225\times 10^{-8}$ \\
$^{25}\mathrm{Mg}$& $-1.65151\times 10^{-2}$ & $1.17959\times 10^{-2}$ & $6.57311\times 10^{-1}$ & $-4.69533\times 10^{-1}$ & $2.83109\times 10^{-2}$ & $-1.12684\times 10^{0}$ & $-2.51234\times 10^{-8}$ \\
$^{27}\mathrm{Al}$& $1.05364\times 10^{-1}$ & $-7.52779\times 10^{-2}$ & $-3.56918\times 10^{0}$ & $2.54884\times 10^{0}$ & $-1.80642\times 10^{-1}$ & $6.11802\times 10^{0}$ & $-2.95252\times 10^{-8}$ \\
$^{29}\mathrm{Si}$& $-6.88009\times 10^{-2}$ & $2.26938\times 10^{-2}$ & $1.47701\times 10^{0}$ & $-4.91593\times 10^{-1}$ & $9.14947\times 10^{-2}$ & $-1.96861\times 10^{0}$ & $-4.64815\times 10^{-8}$ \\
$^{31}\mathrm{P}$ & $4.04528\times 10^{-1}$ & $-1.34221\times 10^{-1}$ & $-3.70087\times 10^{0}$ & $1.23251\times 10^{0}$ & $-5.38749\times 10^{-1}$ & $4.93338\times 10^{0}$ & $-1.09198\times 10^{-7}$ \\

\end{longtable}

\end{document}